\documentclass[aps,prd,twocolumn,showpacs,10pt,superscriptaddress,preprintnumbers,nofootinbib]{revtex4-2}
\usepackage{amsmath,bm}
\usepackage{physics}
\usepackage{graphicx, xcolor}
\usepackage{amssymb}
\usepackage[colorlinks, linkcolor=blue!50!black, citecolor=green!50!black, urlcolor=magenta]{hyperref}
\usepackage[normalem]{ulem}
\usepackage{adjustbox}

\newcommand{\fig}[1]{Fig.~\ref{#1}}
\newcommand{\eq}[1]{Eq.~\eqref{#1}}
\newcommand{\paper}[1]{Ref.~\cite{#1}}
\newcommand*{\Scale}[2][4]{\scalebox{#1}{\ensuremath{#2}}}

\newcommand{\beq}[1][]{\begin{equation}\label{#1}}
\newcommand{\eeq}{\end{equation}}
\newcommand{\bse}[1][]{\begin{subequations}\label{#1}}
\newcommand{\ese}{\end{subequations}}
\newcommand{\nn}{\nonumber}

\newcommand{\pp}[1]{\left(#1\right)}
\newcommand{\bb}[1]{\left[#1\right]}
\newcommand{\cc}[1]{\left\{#1\right\}}

\newcommand{\LQCD}{\Lambda_{\rm QCD}}

\newcommand{\ie}{{\it i.e.}}
\newcommand{\M}{\mathcal{M}}
\newcommand{\e}{\varepsilon}
\newcommand{\hard}{\mathcal{H}}
\newcommand{\G}{{\rm Glauber}}
\newcommand{\D}{{\cal D}}
\newcommand{\mybar}{\rule[-0.55pt]{0.5pt}{7.05pt}}
\newcommand\rt{\sqrt{\mybar \hspace{1.25pt} t \hspace{1.1pt} \mybar}}

\begin{document}

\preprint{
	{\vbox {			
		\hbox{\bf JLAB-THY-22-3742}
		\hbox{\bf MSUHEP-22-032}
}}}
\vspace*{0.2cm}

\title{Single diffractive hard exclusive processes for the study of generalized parton distributions}

\author{Jian-Wei Qiu}
\email{jqiu@jlab.org}
\affiliation{Theory Center, Jefferson Lab,
Newport News, Virginia 23606, USA}
\affiliation{Department of Physics, William \& Mary,
Williamsburg, Virginia 23187, USA}

\author{Zhite Yu}
\email{yuzhite@msu.edu}
\affiliation{Department of Physics and Astronomy, 
Michigan State University, East Lansing, Michigan 48824, USA}

\date{\today}

\begin{abstract}
Generalized parton distributions (GPDs) are important nonperturbative functions that provide tomographic images of partonic structures of hadrons. We introduce a type of exclusive processes, to be referred to as single diffractive hard exclusive processes (SDHEPs).  We discuss the necessary and sufficient conditions for SDHEPs to be factorized into GPDs.  We demonstrate that the SDHEP is not only sufficiently generic to cover all known processes for extracting GPDs, but is also well motivated for the search of new processes for the study of GPDs. We examine the sensitivity of the SDHEP to the parton momentum fraction $x$ dependence of GPDs.
\end{abstract}

\maketitle

\section{Introduction}
\label{sec:intro}

The parton distribution function (PDF), $f_{i/h}(x)$, is a well-studied nonperturbative quantum correlation function in Quantum Chromodynamics (QCD) that describes the distribution of the momentum fraction $x$ of an active parton of flavor $i$ inside a colliding hadron $h$. 
The knowledge of PDFs, in particular, their $x$-dependence, as well as the suppressed factorization scale dependence, is crucial for understanding all phenomena in high energy hadronic collisions, including the Large Hadron Collider, where the colliding hadrons are broken.
The generalized parton distribution functions (GPDs), $F_{i/h}(x, \xi, t)$, 
encode rich information on the spatial distributions of the partons inside a bound hadron, and can be extracted from hard exclusive processes off a diffractive hadron while it is kept intact.
With two additional variables, $\xi$ and $t$, GPDs cover much richer nonperturbative information of the confined partonic dynamics in a hadron, and have been of both theoretical and experimental interest since they were first introduced in 1996~\cite{Muller:1994ses}.
On the one hand, as a combination of the hadronic form factor and PDF, the $x$ moments of GPDs lead to some important sum rules~\cite{Ji:1996ek} and can be related to different form factors of the QCD energy-momentum tensor, providing valuable information on the fundamental properties of hadrons, such as the partonic contribution to the hadron spin~\cite{Ji:1996ek}, the hadron mass~\cite{Ji:1994av, Ji:1995sv, Lorce:2017xzd, Metz:2020vxd}, the pressure and shear force inside a hadron~\cite{Polyakov:2018zvc, Burkert:2018bqq}. 
On the other hand, with the additional scale $t$, the Fourier transform of $F_{i/h}(x, 0, t)$ with respect to the transverse component of $t$ gives a (2+1)-dimensional density distribution, $f_{i/h}(x, \bm{b}_T)$, of the partons inside a confined hadron~\cite{Burkardt:2000za, Burkardt:2002hr}, which provides the hadron's three-dimensional (3D) tomography, entailing a great amount of information on how QCD holds quarks and gluons together to form a bound hadron.
For reviews, see Refs.~\cite{Goeke:2001tz, Diehl:2003ny, Belitsky:2005qn, Boffi:2007yc}.

Along with the introduction of the deeply virtual Compton scattering (DVCS)~\cite{Ji:1996nm, Radyushkin:1997ki}, a number of processes have been proposed for extracting the GPDs from experimental observables~\cite{Brodsky:1994kf, Frankfurt:1995jw, Berger:2001xd, Berger:2001zn, Guidal:2002kt, Boussarie:2016qop, Pedrak:2017cpp, Duplancic:2018bum}. It is the QCD factorization theorem that expresses the physical observables in terms of the convolution of nonperturbative but process-independent GPDs and perturbatively calculable hard coefficients, $F_{i/h}(x, \xi, t) \otimes C_{i}(x, \xi, Q)$, with process-dependent corrections suppressed by powers of the large momentum transfer of the scattering processes. Different processes give different hard coefficients, which act as probes to project out different information of GPDs. The knowledge and understanding of GPDs are obtained when multiple processes are combined in global analyses.
Among the existing processes in the literature, nevertheless, only a few have been strictly proved to be factorizable~\cite{Collins:1996fb, Collins:1998be}. Factorization formalism has been extrapolated to describe other processes while waiting for its formal proof.  

Among the three variables $(x, \xi, t)$ of the GPD, 
both the $\xi$- and $t$-dependence are related to the kinematics of the diffracted hadron, and only the $x$-dependence is probed by the hard {\it partonic} scattering, like the $x$-dependence of PDFs.
However, it is the $x$-dependence of the GPD that is the most difficult to extract from experimental data.  Firstly, due to the amplitude nature of GPDs, the related factorization happens at the amplitude level for {\it exclusive} processes, and the convolution variable $x$ is the parton loop momentum, flowing through the active parton pair defining the GPDs, whose integration is always from $-1$ to $1$, and is never pinned down to a particular value. This is in contrast to the factorization of inclusive processes like the deeply inelastic scattering (DIS), which happens at the cross section level.  At the leading perturbative order, the parton momentum fraction $x$ is equal to the Bjorken-$x_B$, which is a direct experimental observable. Secondly, the GPD-related physical processes rely on an exchange of a color-singlet two-parton state ($q\bar{q}$ or $gg$) between the diffracted hadron we want to study and the ``hard probe", but, the same processes could also happen via the exchange of a virtual photon if its quantum state is allowed. This channel of single photon exchange could dominate the contribution to the total amplitude and interfere with the GPD-sensitive channels, 
causing a large background for extracting the GPDs. Thirdly, for most of the GPD-related processes, the convolutions of the hard coefficients with GPDs only give ``moment-type" information, like the integral $\int_{-1}^1 \dd{x} F(x,\xi,t) / (x \pm \xi \mp i\varepsilon)$ probed by the DVCS.

In a recent paper~\cite{Qiu:2022bpq}, we demonstrated that the last two sources of difficulties for extracting GPDs could be improved by identifying GPD-related physical processes with new types of hard probes.  By considering the exclusive diphoton production process in a diffractive pion-nucleon collision, and requiring the final-state photon's transverse momentum $q_T \gg \rt$, it was shown that the scattering amplitude can be factorized into a nucleon transition GPD, without the interference with a single photon exchange channel, while the variation of $q_T$ provides an extra handle to the $x$-dependence of the GPD. By measuring the distribution of $q_T$ in such GPD-related exclusive process, we can obtain {\it enhanced} sensitivity to the $x$-dependence of GPDs.

\begin{figure}[htbp]
\centering
	\includegraphics[scale=0.6]{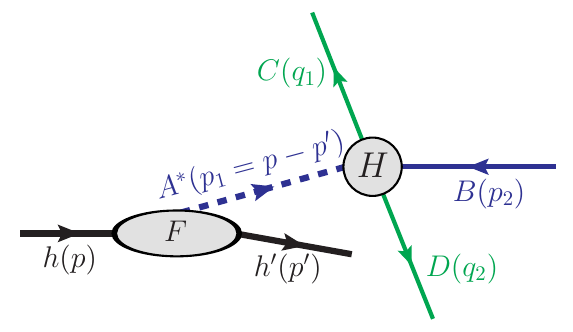}	\\
	(a) \\
	\includegraphics[scale=0.6]{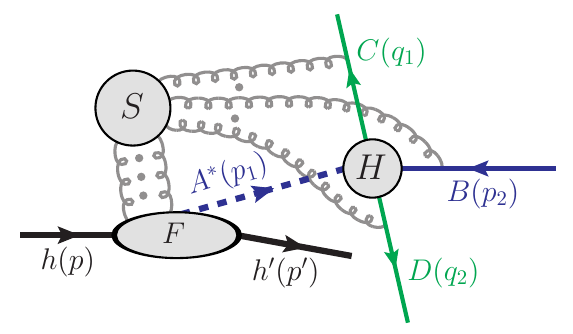} \\
	(b)
\caption{(a) The illustration of the SDHEP as a two-stage reaction. (b) The SDHEP with soft gluon connections in a general case when $B, C$ and $D$ are hadrons.}
\label{fig:sdhep}
\end{figure}

As indicated in Ref.~\cite{Qiu:2022bpq}, the diphoton process can be generalized to, as illustrated in \fig{fig:sdhep}(a),
a generic $2\to3$ {\it single diffractive hard exclusive process} (SDHEP),
\beq
h(p) + B(p_2) \to h'(p') + C(q_1) + D(q_2),
\label{eq:sdp}
\eeq 
where $h$ of momentum $p$ is a hadron we would like to study, $B$ of momentum $p_2$ is a colliding lepton, photon or meson, and $C$ and $D$ of momentum $q_1$ and $q_2$, respectively, are two final-state particles, which can be a lepton, photon or meson, with large transverse momenta, 
\beq[eq:hard qT]
	q_{1T}\sim q_{2T} \gg \rt\, ,
\eeq 
with $t \equiv (p-p')^2$. The SDHEP can be thought of as a two-stage process, being a combination of a diffractive production of a single long-lived state $A^*(p_1)$,
\beq[eq:diffractive]
	h(p) \to A^*(p_1) + h'(p'), \quad \mbox{with } p_1 = p - p',
\eeq
and a hard exclusive $2\to2$ scattering between the two nearly head-on states $A^*(p_1)$ and $B(p_2)$,
\beq[eq:hard 2to2]
	A^*(p_1) + B(p_2) \to C(q_1) + D(q_2).
\eeq
In the center-of-mass (c.m.) frame of $A^*$ and $B$, as a necessary condition for the factorization, the transverse momentum $q_T$ of $C$ or $D$ is required to be much greater than the invariant mass of $A^*$ or $B$. 

In this paper, we will show that the condition in Eq.~\eqref{eq:hard qT} is not only necessary, but also sufficient for the SDHEP to be factorized into hadron GPDs associated with the $h\to h'$ transition, convoluted with perturbatively calculable coefficient functions from the hard $2\to 2$ exclusive scattering, if the final-state particles $C$ and $D$ are produced via a single hard interaction.  We will also demonstrate that the SDHEP is not only sufficiently generic to cover all known processes for extracting GPDs in the literature, but also well-motivated for the search of new processes for the study of GPDs. 

The rest of this paper is organized in the following way.
We begin with a general discussion of the features of the SDHEP and the factorization structure in Sec.~\ref{sec:sdhep}. We then provide the detailed arguments for QCD factorization of SDHEPs, initiated by a lepton, a photon, and a meson beam, respectively, in Secs.~\ref{sec:lepton}-\ref{sec:meson}. 
In Sec.~\ref{sec:discussion}, we will provide an extended discussion on the factorization properties, the limitation of extracting the $x$-dependence of GPDs from some SDHEPs, the strategy to identify SDHEPs that can provide better sensitivity to the $x$-dependence, and additional opportunities for extracting various types of GPDs from SDHEPs. Finally, we provide our summary and outlook in Sec.~\ref{sec:summary}.

\begin{figure*}[htbp]
\def\sc{1.8}
\centering
\begin{align*}
	&\adjincludegraphics[valign=c, scale=0.6]{figures/sdhep.pdf}	
	 \Scale[\sc]{=}
	 \adjincludegraphics[valign=c, scale=0.6]{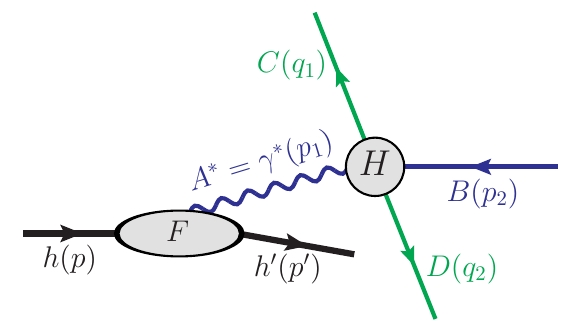} \Scale[\sc]{+} \\
	& \hspace{10em} \Scale[\sc]{+}
	 \adjincludegraphics[valign=c, scale=0.6]{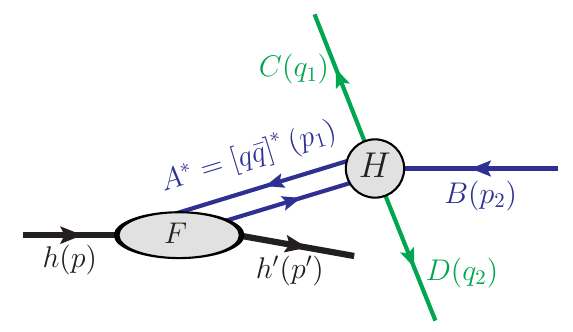} \Scale[\sc]{+} 
	 \adjincludegraphics[valign=c, scale=0.6]{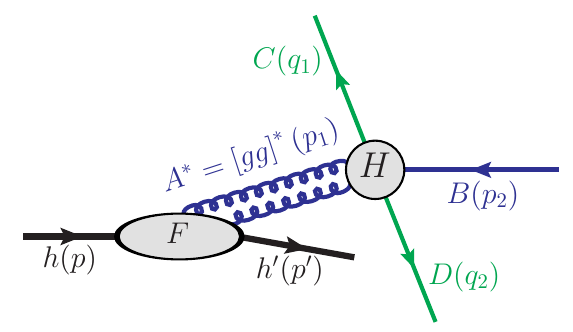} \Scale[\sc]{+\cdots}
\end{align*}
\caption{The representation of the SDHEP in terms of all possible exchanged channels of the virtual state $A^*(p_1)$ between the single diffractive $h\to h'$ transition and the $2\to 2$ hard exclusive process. The two gluons in $gg$ channel have physical polarizations. The $q\bar{q}$ and $gg$ channels can be accompanied by arbitrary numbers of collinear longitudinally polarized gluons. The ``$\cdots$" refers to the channels with more than two physically polarized partons, which are power suppressed compared to the two-parton case.
}
\label{fig:decomposition}
\end{figure*}

\section{Single-diffractive hard exclusive scattering process}
\label{sec:sdhep}

The two-stage paradigm of SDHEP, as illustrated in \fig{fig:sdhep}(a), is necessary to separate the dynamics taking place at two distinct scales $Q (\sim\! q_{1T} \!  \sim\!   q_{2T})$ and $t$. With $Q \gg \rt$, the lifetime of the exchanged state $A^*(p_1)$ at ${\cal O}(1/\rt)$ is so much longer than the timescale $1/Q$ of the ``hard probe'' (the $2\to 2$ hard exclusive subprocess) that the quantum interference mediated by soft gluons between the diffractive hadron and the hard probe, as illustrated in \fig{fig:sdhep}(b), is expected to be strongly suppressed by the power of $\rt/Q$, so that the hard probe is unlikely to alter the internal structure of the hadron that we would like to study.

The $2\to 2$ hard exclusive process $H$ in \fig{fig:sdhep} takes place at a short distance $1/Q \ll 1/\LQCD\sim 1$~fm and is sensitive to the partonic structure of the exchanged state  $A^*(p_1)$. The scattering amplitude of the SDHEP should include a sum of all possible partonic states, as illustrated in \fig{fig:decomposition}, 
which can be schematically described as
\beq[eq:channels]
	\mathcal{M}_{hB\to h'CD} = \sum_{n = 1}^{\infty} \sum_f F_{h\to h'}^{f_n}(p, p') \otimes C_{f_n B \to CD},
\eeq
where $n$ and $f$ represent the number and flavor of particles included in the exchanged state $A^*$, respectively, $F_{h\to h'}^{f_n}(p, p')$ is a ``form factor'' responsible for the $h\to h'$ transition, and $C_{f_n B \to CD}$ represents the scattering amplitude of the hard part $H$, along with the sum running over all possible exchanged states characterized by $n$ and $f$. For the discussion in this paper, we keep the scattering amplitude $C_{f_n B \to CD}$ at the lowest order in the QED coupling constant for given exchanged state $f_{n}$ and particle types of $B$, $C$, and $D$, while we explore contributions from QCD at all orders in its coupling constant.

For $n = 1$, the only possible case is a virtual photon exchange, \ie, $f_1 = \gamma^*$, which is like the Bethe-Heitler process for the DVCS (see Ref.~\cite{Ji:1996nm} for example). Instead of probing the partonic structure of $h$, this channel only gives an access to the electromagnetic form factor of $h$ evaluated at a relatively soft scale $t$.  
As discussed below, the $\gamma^*$-mediated subprocess gives the ``superleading power" background for the $n\geq 2$ channels, and should not be excluded even if they are suppressed by higher orders of QED coupling, unless it is forbidden by the symmetry. The scattering amplitude of the SDHEP should be expanded in inverse powers of the hard scale, and then followed by a perturbative factorization for the leading power contribution (and subleading power contribution if needed, see, e.g., Ref.~\cite{Kang:2014tta}). If the $n=1$ subprocess is forbidden (as discussed below), then the scattering amplitude of the SDHEP starts with $n = 2$ subprocesses.

For $n = 2$, we can have QCD subprocesses with $f_2 = [q\bar{q}']$ or $[gg]$. This gives the leading-power contribution that, as shown in the next section, can be factorized into GPDs with corresponding hard coefficients.  The channels with $n \geq 3$ belong to high-twist subprocesses that are suppressed by powers of $\rt/Q$ and will be neglected in the following analysis.

\begin{figure}[htbp]
\centering
	\includegraphics[scale=0.67]{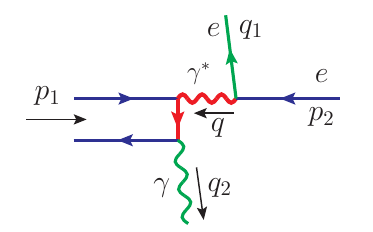}	
	\includegraphics[scale=0.67]{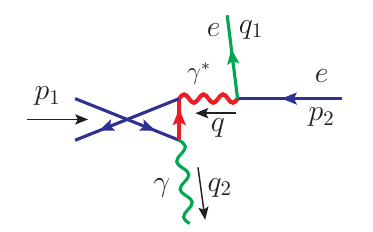}	
\caption{The leading-order diagrams to the hard exclusive subprocess of the DVCS, initialized by the state $f_2 = [q\bar{q}]$. The red thick lines indicate the propagators with high virtualities and thus belong to the hard part.}
\label{fig:dvcs-h}
\end{figure}

The SDHEP in \eq{eq:sdp} is a generalization of the diphoton production process that we considered in \paper{Qiu:2022bpq}, but it is actually general enough to cover all the processes that exist in the literature for extracting GPDs. For example, the DVCS corresponds to $B = C = e$ and $D = \gamma$ with $n = 2$ hard scattering process, $[q\bar{q}](p_1)+e(p_2) \to e(q_1) +\gamma(q_2)$, as shown in Fig.~\ref{fig:dvcs-h}.  The large transverse momentum of the scattered electron $q_{1T}$ (or the $q_{2T}$ of the produced photon) provides the hard scale $Q$. Here we changed the perspective by working in the c.m. frame of the colliding electron and proton, instead of the Breit frame of the virtual photon and colliding proton in the conventional discussion.  By including the scattered electron into the discussion of factorization, which fits into the general SDHEP, it makes more natural to include the Bethe-Heitler process into the observed cross section. Similarly, the deeply virtual meson production (DVMP)~\cite{Brodsky:1994kf, Frankfurt:1995jw} process can be obtained by changing $D$ to a meson. The timelike Compton scattering (TCS)~\cite{Berger:2001xd} corresponds to $B = \gamma$ and $C, D = l^-, l^+$, for which the hard $q_T$ of the leptons provides a hard virtuality $Q \sim q_T$ for the virtual photon which directly couples to the hard interaction, setting the hard scale. The exclusive Drell-Yan process~\cite{Berger:2001zn} is similar by having $B = \pi^{\pm}$. The diphoton photoproduction process~\cite{Pedrak:2017cpp, Grocholski:2021man, Grocholski:2022rqj} corresponds to setting $B = C = D = \gamma$, for which the hard $q_T$ of the final-state photons provides the hard scale. Similarly, the diphoton production in the diffractive pion-nucleon scattering~\cite{Qiu:2022bpq} corresponds to $B = \pi^{\pm}$ and $C = D = \gamma$.  The crossing process of 
photoproduction of a photon-meson pair~\cite{Boussarie:2016qop, Duplancic:2018bum} corresponds to $B = D = \gamma$ and $C=$~meson.

\subsection{General discussion of the $\gamma^*$-mediated channel}
\label{subsec:n=1}

Before providing the detailed arguments for QCD factorization of SDHEPs, initiated by a lepton, photon or meson beam, respectively, in next three sections, we give a general discussion for the $\gamma^*$-mediated hard subprocesses, corresponding to the $n = 1$ channel in \eq{eq:channels}, independent of the particle types of $B, C$ and $D$.  More detailed discussion for specific processes will be given in later sections.

One difference between the $n = 1$ and $n \geq 2$ subprocesses is that the virtual photon momentum is fully determined by the diffraction of the hadron $h$. The amplitude of the $\gamma^*$-mediated subprocess can be trivially factorized into the electromagnetic form factor of the hadron $h$,
\begin{align}\label{eq:n1 FF}
	\M^{(1)} 
	&= \frac{ie^2}{t} \langle h'(p') | J^{\mu}(0) | h(p) \rangle \,
			\langle C(q_1) D(q_2) | J_{\mu}(0) | B(p_2) \rangle	\nn \\
	&\equiv \frac{ie^2}{t} F^{\mu}(p, p') \, \hard_{\mu}(p_1, p_2, q_1, q_2),
\end{align}
where the superscript ``$(1)$'' refers to the contribution to the SDHEP amplitude from the $n = 1$ channel, and $J^{\mu} = \sum_{i \in q} Q_i \bar{\psi}_i \gamma^{\mu} \psi_i$ is the electromagnetic current of quarks, summing over flavor ``$i$'' and weighted by their fractional charges $Q_i$. We defined the form factor,
\begin{align}
	& F^{\mu}(p, p') = \langle h'(p') | J^{\mu}(0) | h(p) \rangle	\nn\\
	& \hspace{1em} 
		= F_1^h(t) \, \bar{u}(p') \gamma^{\mu} u(p) + F_2^h(t) \, \bar{u}(p')\frac{i\sigma^{\mu\nu}p_{1\nu}}{2m_h} u(p),
\end{align}
which has the leading component $F^+ \sim \order{Q}$ as the $h$-$h'$ system is highly boosted along $\hat{z}$ direction.\footnote{We define the light-front components of a vector $V^{\mu} = (V^+, V^-, \bm{V}_T)$ as $V^{\pm} = (V^0 \pm V^3)/\sqrt{2}$ and $\bm{V}_T = (V^1, V^2)$.} However, when this component is contracted with $\hard_{\mu}$, which scales as $\order{Q^0}$ for each component, we have
\begin{align}
	&F^+ \hard^- = \frac{1}{p_1^+} \, F^+ \pp{ p_1^+ \, \hard^-}	\nn\\
	&\hspace{1em}
		= \frac{1}{p_1^+}  \, F^+ \pp{p_1 \cdot \hard + \bm{p}_{1T}\cdot \bm{\hard}_{T} - p_1^- \hard^+ },
\end{align}
where in the bracket, the first term vanishes by the Ward identity of QED, and the other two scale as $\rt$ and $t/p_1^+$ respectively. So the leading power of $F \cdot \hard$ scales as $\rt$ and is given by the transverse polarization of the virtual photon. Therefore, the power counting of $\M^{(1)}$ is of the order $1 / \rt$, which is higher than the $n = 2$ channel by one power of $Q / \rt$. 

One caution should be noted that it is not appropriate to only keep $p_1^+$ in the amplitude $\hard_{\mu}(p_1, p_2, q_1, q_2)$ because the approximation introduces an error of order $\rt/Q$. While this is power suppressed comparing to the leading contribution from the $n = 1$ channel, it could scale at the same order as the contribution from the $n = 2$ channel since both of them have the power counting $1/Q$. By neglecting all the $n \geq 3$ channels, our approximation to the full SDHEP amplitude is up to the error at ${\cal O}(\rt/Q^2)$, so that the $1/Q$ part should be kept as exact when evaluating the contribution from the $n=1$ channel.

We note one further subtlety of the case when the $\gamma^*$-mediated subprocess involves light mesons in $\hard$. 
The conventional practice is to factorize it into meson distribution amplitudes (DAs).
While this is true to the leading power at $\order{1/\rt}$, it neglects the power correction of $\order{\LQCD/Q} \cdot \order{1/\rt} = \order{1/Q}$, which is of the same order as the $n = 2$ channels, \ie, the GPD channels. Keeping the exact $1/Q$ contribution thus requires the subleading-power (or, twist-3) factorization for the $\gamma^*$-mediated subprocess that involves any mesons, which is beyond the scope of this paper.

There are two cases in which the $\gamma^*$-channel is forbidden. The first is for a flavor-changing channel with $h \neq h'$ that cannot be achieved by the electromagnetic interaction, like the pion-nucleon scattering processes in Refs.~\cite{Berger:2001zn, Qiu:2022bpq} which can involve the proton-neutron transition. 
The second case is for particular combinations of the particle types of $B, C$ and $D$ that require $\hard_{\mu}(p_1, p_2, q_1, q_2) = 0$ by some symmetries. 
Apart from these two cases, we should generally include the $\gamma^*$-mediated subprocess. 

For example, for the photoproduction of diphoton process considered in~\paper{Pedrak:2017cpp}, one should include the $\gamma^*$-channel that involves photon-photon scattering in $\hard_{\mu}$. Even though this is suppressed by $\alpha_{\rm em}$ compared to the GPD subprocess that corresponds to the $n = 2$ channel, the $\gamma^*$-channel at $n=1$ is power enhanced by $Q/ \rt$. In such cases, we need to carefully compare the contributions from these two channels,  and to develop an experimental approach to remove the background due to the $\gamma^*$-channel in order to extract GPDs from the experimental data. One common approach by using azimuthal correlations will be briefly discussed in Sec.~\ref{subsec:angular}.

\section{SDHEP with a lepton beam}
\label{sec:lepton}

For single diffractive hard exclusive electroproduction processes, we have $B = C = e$. The other particle $D$ can be a photon $\gamma$ or a light meson $M_D$. Both of these two processes allow the $\gamma^*$-initialized channel with $n = 1$. For the $n = 2$ channel, the $D = \gamma$ case leads to the DVCS process, and the case for $D = $~light meson corresponds to the DVMP process. Both of these two processes have been proved to be factorized into GPDs~\cite{Collins:1998be, Collins:1996fb}. Here, we will switch the theoretical perspective from Refs.~\cite{Collins:1998be, Collins:1996fb} by fitting them into the general SDHEP type of processes. The proof follows the two-stage paradigm depicted in Eqs.~\eqref{eq:diffractive}-\eqref{eq:channels}. This approach incorporates the $\gamma^*$-initialized $n=1$ channel naturally, and for the $n = 2$ channel, it leads to a direct analogy to the exclusive meson annihilation process in \eq{eq:hard 2to2} by having $A^*$ being some meson state carrying the quantum number of the $[q\bar{q}']$ or $[gg]$ state. Our strategy for the proof follows a two-step process introduced in~\paper{Qiu:2022bpq}: (1) justify the factorization for a simpler $2\to 2$ hard exclusive process in \eq{eq:hard 2to2}, and (2) extend the factorization to the full SDHEP in \eq{eq:sdp} by addressing extra complications, including the difficulty from Glauber gluons. As expected, we will reproduce the proofs in Refs.~\cite{Collins:1998be, Collins:1996fb}.

\subsection{Real photon production: $D = \gamma$}
\label{sec:dvcs}

For $n = 1$, this gives the Bethe-Heitler process, and the amplitude $\hard^{\mu}$ in \eq{eq:n1 FF} is the scattering amplitude of $\gamma^*(p_1) + e(p_2) \to e(q_1) + \gamma(q_2)$ with $q_{1T}^2\gg |p_1^2|=|t|$.

\begin{figure}[htbp]
\centering
	\includegraphics[scale=0.7]{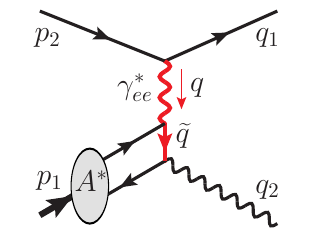}
\caption{Leading-order sample diagram for the DVCS, which occurs by exchanging the virtual state $A^* = [q\bar{q}]$ between the diffractive hadron $h$ and the electron beam.}
\label{fig:dvcs LO}
\end{figure}

\begin{figure}[htbp]
\centering
	\begin{tabular}{cc}
		\includegraphics[scale=0.7]{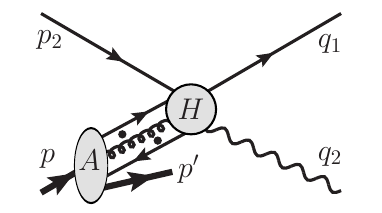} &
		\includegraphics[scale=0.7]{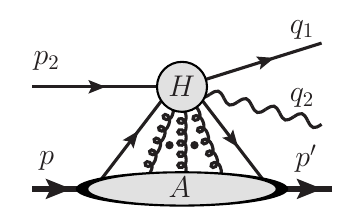}	\\
		(a) & (b) 
	\end{tabular}
\caption{Leading-region graphs of the DVCS for the (a) ERBL region and (b) DGLAP region of the GPD, where the two quark lines can be replaced by two transverse gluon lines.}
\label{fig:dvcs}
\end{figure}

For $n = 2$, the state $A^*$ can be either a collinear $q\bar{q}$ or $gg$ pair, which interacts with the electron beam by exchanging a virtual photon $\gamma_{ee}^*$ with momentum $q = p_2 - q_1$, as shown in Fig.~\ref{fig:dvcs LO}.  The $[q\bar{q}]$ and $[gg]$ state can be accompanied by an arbitrary number of longitudinally polarized collinear gluons. The traditional treatments all work in the Breit frame of the virtual photon $\gamma_{ee}^*$ and hadron beam $h$~\cite{Collins:1998be, Collins:1996fb}. Here, we follow the kinematic setup of the SDHEP in~\eq{eq:sdp} to work in the c.m. frame of the initial-state hadron and electron with the hadron along the $z$ axis.
The requirement of a high virtuality $Q^2 = -q^2$ for the $\gamma_{ee}^*$ is equivalent to the requirement of hard transverse momenta $q_T$ for the final-state electron and photon in this frame, since $Q^2 \propto q_T^2$. Hence, the virtual photon $\gamma_{ee}^*$ has a short lifetime and belongs to the hard part, and therefore we have the leading-region diagrams as in \fig{fig:dvcs}, 
where two leading regions, associated with the ERBL and DGLAP regions, respectively, were identified.  The ERBL region of GPDs corresponds to the region where $|x|<|\xi |$ and GPDs evolve like the evolution of meson DAs~\cite{Efremov:1979qk,Lepage:1980fj}, and the DGLAP region is for $|x| > |\xi |$ where the evolution of GPDs is similar to the DGLAP evolution of PDFs~\cite{Dokshitzer:1977sg,Gribov:1972ri,Lipatov:1974qm,Altarelli:1977zs}.

\subsubsection{The $2\to2$ hard exclusive scattering of SDHEP}
\label{sec:DVCS m}
\begin{figure}[htbp]
\centering
	\includegraphics[scale=0.7]{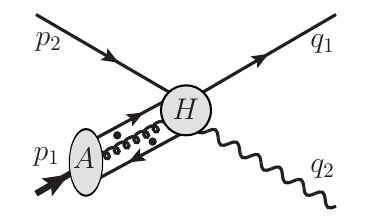}	\\
\caption{Leading-region graphs for the $2\to2$ hard exclusive scattering of real photon electroproduction, which is the hard probe part of the SDHEP with a lepton beam, where the two quark lines from $[q\bar{q}](p_1)$ can be replaced by two transversely polarized gluon lines for the case when $A^*=[gg]$.}
\label{fig:dvcs meson}
\end{figure}

The $2\to2$ hard exclusive scattering of the SDHEP with a lepton beam is effectively the exclusive scattering of an electron and a meson into an electron and a photon,
\beq[eq:dvcs m]
	M_A(p_1) + e(p_2) \to e(q_1) + \gamma(q_2).
\eeq
In the c.m. frame with the meson going along $\hat{z}$ direction, the final-state electron and photon are required to have high transverse momenta. One leading-order diagram has been shown in \fig{fig:dvcs LO}, where the red thick lines carry the hard $q_T$ flow and so have high virtualities and belong to the hard part. The general leading-region diagram is in \fig{fig:dvcs meson}, where the meson-collinear $q\bar{q}$ lines can also be replaced by a pair of transversely polarized gluon lines, both of which can be accompanied by an arbitrary number of collinear gluons whose polarizations are proportional to their momenta.

Perturbatively, factorization is to organize the loop momentum integrals to factorize the pinch singularities from all partonic scattering diagrams into universal hadronic functions, leaving the remaining partonic scattering contribution to be infrared-safe hard coefficient functions. In the simple case in \fig{fig:dvcs meson}, pinch singularity occurs for each parton momentum $k_i$ at small $k_i^-$ as $\lambda = \sqrt{p_1^2} / Q \to 0$, where $Q\sim q_T \sim p_1^+$. 
In QCD with confinement at low energy scale, the perturbative pinch singularity factorized from partonic scattering diagrams is effectively removed by nonperturbative QCD effects once absorbed into the universal hadronic functions. Therefore, factorization is to consistently separate the infrared-sensitive nonperturbative physics from the short-distance hard dynamics, which is free from low energy scale dependence and is purely perturbative.

The factorization of the leading region in \fig{fig:dvcs meson} can be carried out straightforwardly, following the same argument as \paper{Qiu:2022bpq}. In the region around the pinched poles, each of the parton momenta scales as 
\beq[eq:col scaling]
	k_i^{\mu} = \pp{k_i^+, k_i^-, \bm{k}_{iT} } \sim \pp{1, \lambda^2, \lambda} Q.
\eeq
Then we can do the following approximations up to the error of $\order{\lambda}$:
\begin{enumerate}
\item[(1)] 
	keep only the plus components for the collinear momenta flowing into $H$, 
\item[(2)] 
	insert spinor or Lorentz projectors to the collinear quark lines or transversely polarized gluon lines, and 
\item[(3)] 
	keep only the leading Lorentz components for the longitudinally polarized gluons coupling to $H$.
\end{enumerate}
This factorizes the collinear subgraph $A$ from the hard part $H$. Especially, the approximation (3) allows the use of Ward identity to decouple all the longitudinally polarized gluons from the hard part and attach them to two gauge links, one for each of the $q$ and $\bar{q}$ or each of the two transversely polarized gluon lines, for the $[q\bar{q}]$ or $[gg]$ case, respectively. As a result, the subgraph $A$ becomes a standardly defined meson DA 
and only convolutes with $H$ via a plus momentum flow $k^+ = z \, p_1^+$, and the exclusive scattering amplitude takes a factorized form
\beq[eq:m1factorize]
	\M_{M_A e\to e\gamma} = \sum_{i} \int_0^1 \dd{z} \phi_{i/A}(z) \, C_{ie \to e\gamma} (z; q_T),
\eeq
which is valid up to an error of $\order{\lambda}$.  In Eq.~(\ref{eq:m1factorize}), $\phi_{i/A}(z)$ is the meson DA, the sum over $i$ runs over the parton flavors, $[q\bar{q}]$ and $[gg]$, as well as their spin structures, and the hard coefficient $C_{ie \to e\gamma}$ is the scattering between the electron and an on-shell, color-neutral and collinear $q\bar{q}$ or $gg$ pair. We refer to \paper{Qiu:2022bpq} for more details of the derivation.

It is worth emphasizing that the above approximation is true only for the scaling in \eq{eq:col scaling}, which corresponds to the pinch surface whose surrounding region gives the leading-power contribution to the amplitude. In principle, one should keep the scaling $k_i^+ \sim \order{Q}$ throughout the factorization analysis. Nevertheless, in the result of factorization, \eq{eq:m1factorize}, the variable $z$ is integrated from $0$ to $1$, which means that we have to include the region where one of the active partons has momentum $k_i^+ \ll Q$. Perturbatively, this does not lead to a pinch, so we should have deformed the contour of $k_i^+$ by $\order{Q}$ to make the associated propagator in the hard subgraph to have high virtuality. For example, the leading-order hard coefficient contains a term that is proportional to $1/(z - i\varepsilon)Q^2$ which becomes soft as $z\to 0$, and we should deform the contour of $z$ to the lower half complex plane to make $\Im z \sim \order{1}$. Similar issue arises as $z\to 1$. However, since the DA only has support in $z\in[0,1]$, such deformation is forbidden by the end points of the $z$ integration. Therefore, the validity of the DA factorization in \eq{eq:m1factorize} needs to be supplemented with an additional assumption that the end point region should be strongly suppressed by the DA, which we refer to as the soft-end suppression. This situation could be improved by the Sudakov suppression factor introduced in \paper{Li:1992nu}. We hope to come back to this issue in the future.

\begin{figure}[htbp]
\centering
	\includegraphics[scale=0.7]{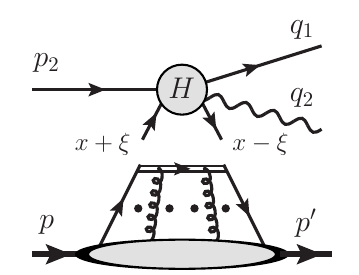}
\caption{Factorization of the DVCS amplitude.}
\label{fig:dvcs factorize}
\end{figure}

\subsubsection{DVCS from SDHEP with a lepton beam}
\label{sec:DVCS p}
Going from the $2\to2$ hard exclusive electron-meson scattering to the full SDHEP with a lepton beam introduces the complication that we have two kinds of leading regions, as shown in \fig{fig:dvcs}, with \fig{fig:dvcs}(a) corresponding to the ERBL region of the GPD that has all the collinear parton lines going from the hadron into the hard interaction, and \fig{fig:dvcs}(b) corresponding to the DGLAP region of the GPD that has some parton lines coming out of the hard interaction to combine with the spectator lines in the beam to form the diffracted hadron $h'$. The extra DGLAP region is the key difference for factorization between the electron-meson scattering and the full SDHEP.  

In the presence of another collinear region, soft gluons exchanged between the spectators of the DGLAP region and the lines from other collinear subgraphs are partially pinched in the Glauber region, for which extra treatment is needed~\cite{Collins:1996fb, Qiu:2022bpq}. However, the DVCS only has one collinear subgraph, and hence its factorization does not have much difference from the electron-meson scattering discussed above.

For both the ERBL and DGLAP regions, the collinear momenta $k_i^{\mu}$ are pinched for their minus components if $\rt  \ll p_1^+ \sim q_T$. Introducing the new scaling variable $\lambda =\rt \, / q_T \ll 1$, the collinear momentum scaling is the same as in \eq{eq:col scaling}. And then the same approximations can be made to factorize the collinear subgraph from the hard subgraph, and lead to a factorization formula for the DVCS amplitude
\beq
	\M^{(2)}_{he \to h' e\gamma} = \sum_i \int_{-1}^1 \dd{x} F^h_i(x, \xi, t) \, C_{ie \to e\gamma}(x, \xi; q_T),
	\label{eq:dvcs factorize}
\eeq
up to terms suppressed by powers of $\lambda$, where the superscript ``(2)'' refers to the contribution to the SDHEP amplitude from the $n = 2$ channel,
$F^h_i$ denotes the flavor-diagonal GPD of the hadron $h$, with $h' = h$, $C_{ie \to e\gamma}$ is the corresponding hard coefficient and $i$ denotes different parton flavors as well as different spin structures. From the fact that the DVCS amplitude should not depend on how we factorize it, renormalization group improvement for the factorization formula in \eq{eq:dvcs factorize} leads to the evolution equations of GPDs with respect to factorization scale $\mu$ and the corresponding $\mu$ dependence in the perturbatively calculated hard coefficient functions, which has been suppressed in \eq{eq:dvcs factorize}. 
Equation \eqref{eq:dvcs factorize} is diagrammatically shown in \fig{fig:dvcs factorize}, in which the GPD variables $x$ and $\xi$ are defined as
\beq
	x = \frac{(k + k')^+}{(p + p')^+}, \quad \xi = \frac{(p - p')^+}{(p + p')^+},
\eeq
where $k$ and $k'$ are the parton momenta entering and leaving the hard part, respectively. 
With the conventional definitions,
\beq
	P = (p + p') / 2,
	\quad
	\Delta = p - p',
\eeq
we have $k^+ = (x + \xi) P^+$, and $k^{\prime + } = (x - \xi) P^+$.

The soft parton issue can also arise here, similar to the electron-meson process discussed at the end of Sec.~\ref{sec:DVCS m}, \ie, some of the parton momenta may have $k_i^+ \ll Q$, which violates the scaling in \eq{eq:col scaling}, and thus the corresponding approximations. This is termed the ``breakpoint" issue in~\paper{Collins:1996fb}. However, since the region $k_i^+ \sim 0 \ll Q$ is not pinched, we can deform the contour of $k^+$ integration by $k^+ \mapsto k^+ \pm i \order{Q}$~\cite{Collins:1996fb}. Because the breakpoint only lies on the boundary between the ERBL and DGLAP regions, but not at the GPD end points, this deformation is allowed. 
Perturbatively, the soft parton singularity appears in \eq{eq:dvcs factorize} at $x = \pm \xi$. For example, the leading-order DVCS hard coefficient contains a term that is proportional to $1/(x\pm \xi \mp i\varepsilon)$, and we can deform the $x$ contour to avoid the poles at $\mp \xi$; in practical calculations, this is achieved by 
\beq[eq:principle value]
	\frac{1}{x \pm \xi \mp i \varepsilon} = P \frac{1}{x \pm \xi} \pm i \pi \, \delta(x \pm \xi),
\eeq
where $P$ denotes principal-value integration.

\begin{figure}[htbp]
\centering
	\begin{tabular}{cc}
		\includegraphics[scale=0.7]{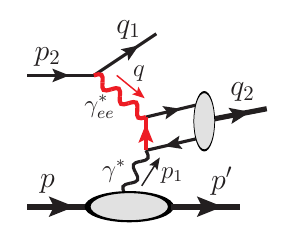} &
		\includegraphics[scale=0.7]{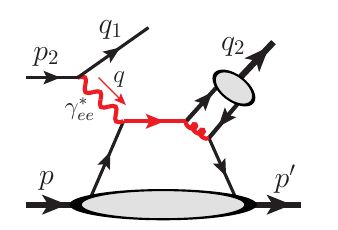}	\\
		(a) & (b) 
	\end{tabular}
\caption{Examples of leading-order diagrams for the light meson production in the SDHEP with an electron beam, for (a) the $n = 1$ channel and (b) the $n = 2$ channel for $[q\bar{q}']$ case, where the red thick lines indicate those with hard $q_T$ flow and high virtualities.}
\label{fig:dvmp lo}
\end{figure}

\subsection{Light meson production: $D=$~meson}
\label{sec:e M}
For the production of a light meson, we have $D=$~meson, with its mass $m_D$ being much smaller than $q_T$. Then the meson is attached to the hard part by a few collinear parton lines, whose momenta are pinched at low virtuality.

The $n = 1$ subprocess is shown in \fig{fig:dvmp lo}(a), and the amplitude $\hard^{\mu}$ in \eq{eq:n1 FF} is the scattering amplitude of $\gamma^*(p_1) + e(p_2) \to e(q_1) + M_D(q_2)$, which can be further factorized into the DA of the meson, up to corrections of order $m_D / q_T$. 
Alternatively, one may choose to parametrize the amplitude by the $\gamma^*\gamma^*_{ee} \to M_D$ form factor.
The $n = 1$ channel would be forbidden for the production of a charged meson like $\pi^{\pm}$, or of a neutral meson with odd $C$ parity, such as $\rho$ and $J/\psi$.

The $n = 2$ subprocess corresponds to the DVMP process~\cite{Brodsky:1994kf, Frankfurt:1995jw}. The hard interaction happens between a $[q\bar{q}']$ or $[gg]$ pair and the colliding electron by exchanging a virtual photon $\gamma_{ee}^*$ and producing a light meson. 
Similar to the DVCS, we work in the $h$-$e$ c.m. frame, and the virtual photon $\gamma^*_{ee}$ belongs to the hard part. One leading-order diagram is shown in \fig{fig:dvmp lo}(b) for $A^* = [q\bar{q}']$ channel. Most part of the factorization follows the same line as for DVCS, and we will only focus on the difference.

\begin{figure}[htbp]
\centering
	\begin{tabular}{cc}
		\includegraphics[scale=0.7]{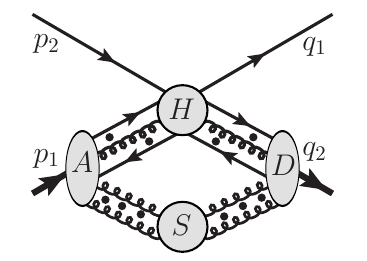} &
		\includegraphics[scale=0.7]{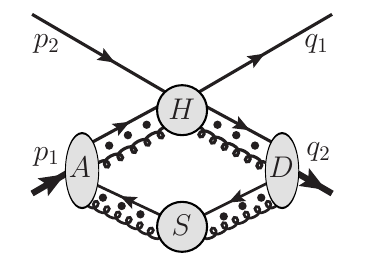}	\\
		(a) & (b) 
	\end{tabular}
\caption{Leading-region graphs for the $2\to2$ hard exclusive electroproduction of a light meson. Depending on the quantum numbers, the quark lines may be replaced by transversely polarized gluon lines.}
\label{fig:m2 leading}
\end{figure}

\subsubsection{The $2\to2$ hard exclusive scattering of SDHEP}
\label{sec:DVMP m}
The $2\to2$ hard exclusive scattering for producing a light meson from SDHEP with a lepton beam is given by
\beq[eq:DVMP m]
	M_A(p_1) + e(p_2) \to e(q_1) + M_D(q_2),
\eeq 
where $M_A$ is a meson state capturing the quantum number of the state $A^*$, which is $[q\bar{q}']$ or $[gg]$. In the c.m. frame, the initial-state $M_A$ and $e$ are along $\pm\hat{z}$ directions, respectively, and the final state $e$ and $M_D$ are back to back, with a transverse momentum much greater than their mass, $q_T \gg m_{A}, m_{D}$. By the same power counting as in  \paper{Qiu:2022bpq}, the leading region of the amplitude is shown in \fig{fig:m2 leading}.

The differences from the real photon production process discussed in Sec.~\ref{sec:DVCS m} are that
\begin{enumerate}
\item[(1)]
there are two collinear subgraphs now, which are connected by a soft subgraph, and
\item[(2)]
there are two kinds of leading regions, shown in Figs.~\ref{fig:m2 leading}(a) and~\ref{fig:m2 leading}(b), which we denote as region (a) and region (b). For region (b), only one active parton enters the hard interaction, and the other one is soft and only transmits the needed quantum number.
\end{enumerate}
Region (b) raises some theoretical difficulty. It is rendered power suppressed in \paper{Collins:1996fb} by considering the scaling $k_s \sim (\lambda, \lambda, \lambda) \, Q$ to be the {\it genuine} soft momentum.
In our case, we simply note that in region (b), the outgoing meson $D$ only has one active parton carrying all of the light-cone momentum, which we assume to be highly suppressed when the meson moves fast ($q_T \gg \LQCD$)~\cite{Qiu:2022bpq}; this assumption is the same as the 
soft-end suppression assumption made in \eq{eq:m1factorize} as discussed at the end of Sec.~\ref{sec:DVCS m}.\footnote{This assumption only holds for the meson side, but not for the diffractive hadron side, and thus does not apply to the backward scattering processes considered in Ref.~\cite{Pire:2021hbl}, for which more theoretical efforts are needed to deal with the region (b).} 
This brings the leading regions down to the one in Fig.~\ref{fig:m2 leading}(a).

To simplify the following discussion, we note that by virtue of the large $q_T$, one can always boost to the frame where the initial-state meson $A^{(*)}$ is moving along $+\hat{z}$ direction and the final-state meson $D$ is moving along $-\hat{z}$ direction, 
as was done in Refs.~\cite{Collins:1981ta, Nayak:2005rt}. 
This can also be done in a covariant way by defining two sets of light-cone vectors
\begin{align}\label{eq:dvmp aux vectors}
	&w_A^{\mu} = \frac{1}{\sqrt{2}}\pp{1, \hat{z} },
	&&\bar{w}_A^{\mu} = \frac{1}{\sqrt{2}}\pp{1, -\hat{z} },
	\nn\\
	&w_D^{\mu} = \frac{1}{\sqrt{2}}\pp{1, \hat{n} },
	&&\bar{w}_D^{\mu} = \frac{1}{\sqrt{2}}\pp{1, -\hat{n} },
\end{align}
where $\hat{z}$ and $\hat{n}$ are normalized three-vectors along the directions of the initial-state meson $M_A$ and final-state meson $M_D$.
Then any momentum four-vector $r$ can be expanded in the $w_A$-$w_D$ frame as
\beq[eq:frame]
	r^{\mu} = r^+ \, w_A^{\mu} + r^- \, w_D^{\mu} + r_{T}^{\mu},
\eeq
where $r^{\pm} = ( r \cdot w_{D,A} ) / ( w_A\cdot w_D )$ are the longitudinal components, and $w_A \cdot w_D \sim \order{1}$ does not affect the power counting. Under this notation, we have
\beq
	r^2 = 2 \, r^+ r^- w_A \cdot w_D - \bm{r}_{T}^2,
\eeq
where $\bm{r}_{T}^2 = - g_{\mu\nu} r_{T}^{\mu} r_{T}^{\nu}$. The $A$-collinear momentum $k_A$ and $D$-collinear momentum $k_D$ have dominant components along $w_A$ and $w_D$, respectively,
\begin{align}\label{eq:cov scaling}
	k_A^{\mu} &= \pp{k_A^+, k_A^-, \bm{k}_{A,T}}_{A\mbox{-}D} \sim (1, \lambda^2, \lambda) Q,	\nn\\
	k_D^{\mu} &= \pp{k_D^+, k_D^-, \bm{k}_{D,T}}_{A\mbox{-}D} \sim (\lambda^2, 1, \lambda) Q,
\end{align}
where the subscript ``$A\mbox{-}D$" refers to light-front coordinates in the $w_A$-$w_D$ frame. In the following discussion of this subsection, we will stay in this frame and omit the subscripts ``$A\mbox{-}D$".

Collinear factorization means that only the longitudinal parton momentum components are seen by the rest of the scattering system. From the point of view of the soft gluons, they only see the large light-cone momenta of the collinear partons but not their transverse momenta. This is true for the soft momentum of a uniform scaling 
\beq
	k_s\sim (\lambda_s, \lambda_s, \lambda_s) Q,
\label{eq:soft k}
\eeq
with $\lambda_s$ varying between $\lambda^2$ and $\lambda$, but not for the Glauber scaling that is dominated by the transverse component,
\beq[eq:Glauber]
	k_s^{\G} \sim (\lambda^2, \lambda^2, \lambda)Q,
\eeq 
which is also part of the soft momentum region and gives a leading-power contribution. However, since all the collinear parton lines connecting $A$ to $H$ move from the past into $H$ with positive plus momenta, and all the collinear parton lines connecting $D$ to $H$ move from $H$ to the future into $D$ with positive minus momenta, the only soft poles for $k_s^+$ ($k_s^-$) come from the $D$- ($A$-) collinear lines, which all lie on the same side of the integration contour in the complex plane, so that the Glauber poles are not pinched. For a Glauber gluon momentum $k_s$ flowing from $A$ into $S$, we deform the $k_s^-$ contour by $k_s^- \mapsto k_s^- - i \order{\lambda Q}$, and for a Glauber gluon momentum $k_s$ flowing from $D$ into $S$, we deform the $k_s^+$ contour by $k_s^+ \mapsto k_s^+ + i \order{\lambda Q}$. This leads the $k_s$ contour out of Glauber region and brings $k_s^{\pm}$ to the same order of $k_{sT}$, after which one may keep only the minus (plus) component of a soft gluon momentum flowing along the $A$- ($D$-) collinear lines. 

After the deformation of soft gluons out of the Glauber region, one can perform suitable approximations for the factorization argument~\cite{Qiu:2022bpq}, especially for the use of Ward identities. The approximations should not introduce extra poles that forbid the above deformations, so if we are looking at a soft gluon of momentum $k_{sA}$ flowing from $A$ into $S$, we should have $1 / (k_{sA}^- - i\varepsilon)$ if the approximator has $k_{sA}^-$ in the denominator. Similarly, for a soft gluon $k_{sD}$ flowing from $D$ into $S$, we should have $1 / (k_{sD}^+ + i\varepsilon)$. The result is that the soft gluons are captured by two sets of Wilson lines, one along the $A$-collinear direction from infinite past to now, and the other along the $D$-collinear direction from now to infinite future. Also, when we make approximations for a collinear gluon, the same gluon momentum $k_c$ can also reach into the soft region (which needs to be subtracted to avoid double counting), so the approximation should not introduce any additional pole at $k_c = 0$. Therefore, for the collinear momentum $k_A$ to flow from $A$ into $H$, we should have $1 / (k_A^+ - i\varepsilon)$ if the approximator has $k_{A}^+$ in the denominator. Similarly, for a collinear momentum $k_D$ to flow from $H$ into $D$, we should have $1 / (k_D^- - i\varepsilon)$. The result is that the collinear-to-$A$ gluons are collected by two Wilson lines along $\bar{w}_1$, pointing to the future, and the collinear-to-$D$ gluons are collected by two Wilson lines along $\bar{w}_D$, pointing to the past.

The above deformation to get $k_s$ contour out of the Glauber region is symmetric with $k_s^+$ and $k_s^-$, as was employed in \paper{Qiu:2022bpq}. This is, nevertheless, not the unique choice~\cite{Collins:2004nx}, as it is sufficient to get rid of the Glauber region as long as $\abs{ k_s^+ k_s^- } \gtrsim \abs{ k_{sT}^2 }$. By examining the contour of $k_s^+$, we note that while all the $k_s^+$ poles from the $D$-collinear lines are of $\order{\lambda^2 Q}$ and lie on the same half plane, it also has poles from the $A$-collinear lines and soft lines, which are of order $Q$ for Glauber gluons. Hence one may choose to only deform the contour of $k_s^+$, but now by a magnitude of $\order{Q}$,
\beq[eq:ks+ deform]
	k_s^+ \mapsto k_s^+ + i \order{Q},
\eeq
when $k_s$ flows from $D$ into $S$. This deforms a Glauber gluon momentum into the $A$-collinear region with the scaling $(1, \lambda^2, \lambda) Q$, and then one can perform usual approximations and apply Ward identities for the rest of the soft gluon momenta. 
The soft gluons factorized from $D$ are attached to two Wilson lines along $w_D$, and the $A$-collinear longitudinally polarized gluons are collected by two Wilson lines along $\bar{w}_A$; both of the two sets of Wilson lines point to the future.
Since we do not deform the contour of $k_s^-$, it does not matter what $i\varepsilon$ prescription we assign to the approximator $1/k_s^-$; the $+i\varepsilon$ choice leads to same result\footnote{Here $k_s$ is the same as in \eq{eq:ks+ deform}, flowing from $D$ to $S$ and then to $A$.} as the symmetric deformation in the above, with soft Wilson lines along $w_A$ and collinear Wilson lines along $\bar{w}_D$ both pointing from/to the past, but the $-i\varepsilon$ choice would have both point to the future.

Similarly, one may also choose to only deform $k_s^-$ as $k_s^- \mapsto k_s^- - i\order{Q}$ when it flows out of $A$-collinear lines into $S$, and then the $i\varepsilon$ prescription for $k^+$ is not important as long as every $k_s^-$ is associated with the same prescription as in $1/(k_s^- + i\varepsilon)$. 

This gives some freedom in choosing the suitable $i\varepsilon$ prescriptions to achieve universal definitions for the soft factor and collinear factors when compared with other processes~\cite{Collins:2004nx}. Within collinear factorization framework, the soft factor cancels no matter what prescription is used, and the Wilson lines associated with the collinear factors also become straight lines on the light cone due to unitarity of the Wilson lines, so that universality is a trivial property in the collinear factorization for exclusive processes. However, such freedom as in \eq{eq:ks+ deform} is necessary for the factorization of diffractive processes, as we will discuss later.

The factorization and cancellation of the soft gluons follow the same procedure as detailed in Ref.~\cite{Qiu:2022bpq} and will not be reproduced here. Physically speaking, with only $k_s^+$ or $k_s^-$ kept in the collinear subgraphs, the soft gluons only see the directions of the collinear lines and not their interior transverse structures. Along each collinear direction defined by the high-momentum hadrons, the collinear parton lines altogether form a color singlet state, so the soft gluons are effectively attached to a color-neutral object, and hence must cancel. This is the special feature of exclusive processes, and is different from the inclusive processes where the soft interaction effectively sums to unity as a consequence of unitarity~\cite{Collins:1981ta, Collins:1988ig}.

After the cancellation of soft gluons, different collinear subgraphs are dynamically independent, and the collinear lines can be factorized into universal meson DAs following the same method as for the photon production case in Sec.~\ref{sec:DVCS m}, leading to a factorized form for the amplitude of the $2\to 2$ scattering in \eq{eq:DVMP m},
\begin{align}\label{eq:m2factorize}
	\M_{M_A e\to e M_D} = &\, \sum_{i, j} \int_0^1 \dd{z_A} \dd{z_D} \phi_{i/A}(z_A) \nn\\
		& \times C_{i e \to e j} (z_A, z_D; q_T) \, \phi_{j/D}(z_D),
\end{align}
where $\phi_{i/A}(z_A)$ and $\phi_{j/D}(z_D)$ are the DAs associated with the initial-state $M_A$ and final-state $M_D$ respectively, $i$ and $j$ run over all possible parton channels, $[q\bar{q}']$ or $[gg]$, and their spin structures, and the hard coefficient $C_{i e \to e j}$ is the scattering between an electron and parton pair $i$ into an electron and parton pair $j$.

\begin{figure}[htbp]
\centering
	\begin{tabular}{cc}
		\includegraphics[scale=0.65]{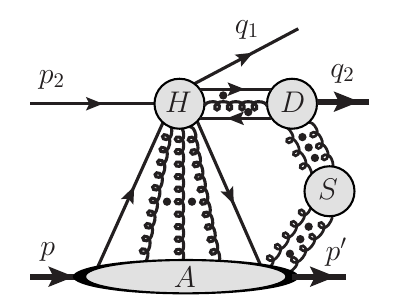} &
		\includegraphics[scale=0.65]{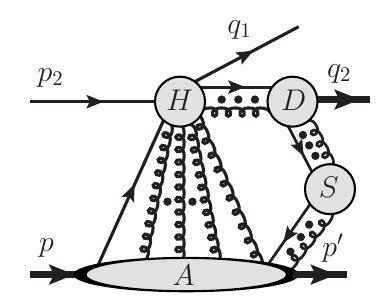}	\\
		(a) & (b) 
	\end{tabular}
\caption{Leading-region graphs for producing a light meson from the SDHEP with a lepton beam. Depending on the quantum numbers, the quark lines may be replaced by transversely polarized gluon lines.}
\label{fig:p2 leading}
\end{figure}

\subsubsection{DVMP from SDHEP with a lepton beam}
\label{sec:DVMP full}
Extending the factorization of $2\to2$ hard exclusive meson scattering, discussed in Sec.~\ref{sec:DVMP m}, to the same meson production from the full SDHEP with a lepton beam introduces complications from the DGLAP region of GPDs, as explained at the beginning of Sec.~\ref{sec:DVCS p}. Factorization works in the limit $\lambda = \rt / q_T \sim m_D / q_T \ll 1$. The loop momentum regions contributing to the leading power of $\lambda$ are shown in \fig{fig:p2 leading}, where the region (b) is rendered power suppressed by the soft-end suppression assumption from the meson wave function, in the same way as for the $2\to2$ meson scattering discussed in Sec.~\ref{sec:DVMP m}. 

\begin{figure*}[htbp]
\centering
	\begin{tabular}{ccc}
		\includegraphics[scale=0.65]{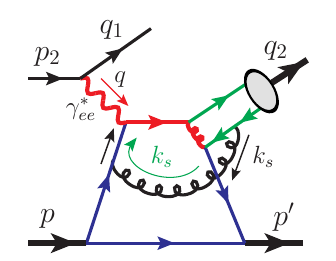} &
		\includegraphics[scale=0.65]{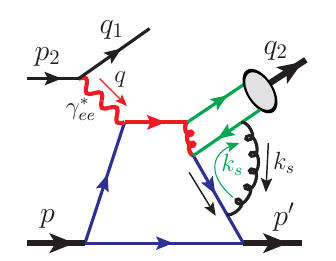} &
		\includegraphics[scale=0.65]{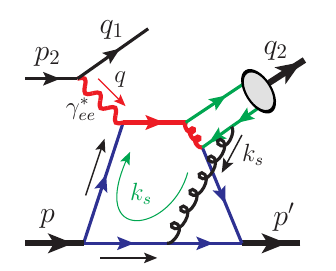}	\\
		(a) & (b) & (c)
	\end{tabular}
\caption{Three example diagrams illustrating the soft gluon exchange between the collinear subgraphs along the diffractive hadron and the final-state meson, for the DGLAP region of the GPD in a simple model theory. The green thin curved lines indicate the soft momentum flows.}
\label{fig:dvmp glauber}
\end{figure*}

While the DGLAP region does not cause much difference for the DVCS, it does lead to a complication in the Glauber region here. This is illustrated in a simple model theory in \fig{fig:dvmp glauber}, where we have indicated the chosen soft momentum flows by the thin curved arrowed lines. We make the following observations:
\begin{enumerate}
\item[(1)]
	DGLAP region has active collinear parton lines both before and after the hard interactions, and the soft gluons can attach to both, as shown in Figs.~\ref{fig:dvmp glauber}(a) and~\ref{fig:dvmp glauber}(b). With the soft momentum flows as indicated, attaching to the initial-state collinear parton gives a pole of $k_s^-$ at $\order{\lambda^2} Q - i\e$, while the final-state one gives a pole of $k_s^-$ at $\order{\lambda^2} Q + i\e$;
\item[(2)]
	DGLAP region also has some spectator partons going in the forward direction. When the soft gluon attaches to the spectator lines, as shown in \fig{fig:dvmp glauber}(c), it flows both in the same and opposite directions as the target-collinear lines, so that one single diagram gives both $\order{\lambda^2} Q \pm i\e$ poles for $k_s^-$ contour.\footnote{Rerouting the soft momentum flow can change the situation (1) such that it also flows through the spectators and leads to both kinds of poles.}
\end{enumerate}
Diagrams like \fig{fig:dvmp glauber}(c) pinch the $k_s^-$ contour at small values, such that for a Glauber gluon with the momentum scaling as in \eq{eq:Glauber}, one cannot deform the $k_s^-$ contour to get out of the Glauber region, as was allowed by the corresponding $2\to2$ scattering. 
However, all the soft $k_s^+$ poles come from the $D$-collinear lines, and lie on the lower half plane when $k_s$ flows from $D$ into $S$.
One may thus deform $k_s^+$ as $k_s^+ \mapsto k_s^+ + i \order{Q}$ while keeping $k_s^-$ contour unchanged, as was done in \eq{eq:ks+ deform}.
While it is a free choice for the $2\to2$ hard exclusive scattering, this deformation is necessary here due to the pinch in the DGLAP region of the diffractive process, and it moves all Glauber gluon momenta to the $A$-collinear region. 
For the same reason as discussed around \eq{eq:ks+ deform}, the $i\e$ prescription for $k^-$ does not matter so it can be chosen in an arbitrary but consistent way.

\begin{figure}[htbp]
\centering
	\includegraphics[scale=0.7]{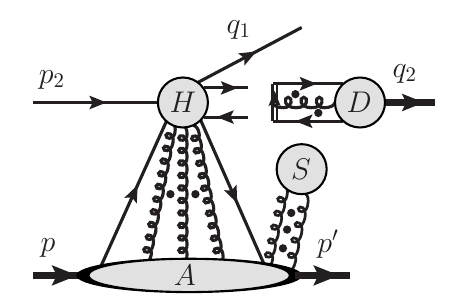} 
\caption{Factorization of soft gluons from the final-state meson for the meson production in the SDHEP with an electron beam.}
\label{fig:p2 soft}
\end{figure}

\begin{figure}[htbp]
\centering
	\includegraphics[scale=0.7]{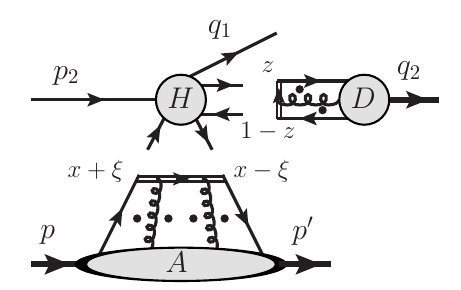}
\caption{Factorization for the the meson production in the SDHEP with an electron beam.}
\label{fig:p2 factorize}
\end{figure}

Our treatment here follows the method in \paper{Qiu:2022bpq}. After deforming the contour of $k^+$ to get rid of the Glauber region, one can apply collinear approximations for $D$ and soft approximations for the gluons coupling $S$ to $D$. This step decouples soft gluons from $D$ and $D$-collinear lines from $H$, the details of which are referred to \paper{Qiu:2022bpq}. After cancellation of the soft gluons attached to $D$, the rest of the soft gluons only couple to the $A$ subgraph, as shown in Fig.~\ref{fig:p2 soft}, which are not pinched~\cite{Collins:1998be} and can be deformed into the $A$-collinear region. By using the same collinear approximations, we can factorize the $A$-subgraph into universal GPDs, and arrive at a factorized amplitude,
\begin{align}\label{eq:p2 factorize}
	\mathcal{M}^{(2)}_{he \to h' eM_D} = &\, \sum_{i, j} \int_{-1}^1 \dd{x} \int_0^1 \dd{z_D} F^{hh'}_i(x, \xi, t)	\\
	&\times C_{ie \to ej}(x, \xi; z_D; q_T) \, \phi_{j/D}(z_D),	\nn
\end{align}
up to $1/q_T$ power suppressed terms.  In Eq.~(\ref{eq:p2 factorize}), the notations are the same as in \eq{eq:m2factorize}, and $F^{hh'}_i$ is the GPD associated with the $h\to h'$ transition. The hard coefficient $C_{ie \to ej}(x, \xi; z_D; q_T)$ can be calculated in the same way as that in \eq{eq:m2factorize}, just with a proper variable change $z_A \to z_A(x, \xi) = (x + \xi)/(2\, \xi)$, which was made in \paper{Qiu:2022bpq}. \eq{eq:p2 factorize} is diagrammatically shown in \fig{fig:p2 factorize}.

\subsection{Virtual photon or heavy quarkonium production}
\label{subsec:e virtual photon}

The DVCS and DVMP differ in how the observed particle couples to the hard interaction: the photon of the DVCS couples directly to the hard collision while the light meson of DVMP couples to the hard collision via two collinear partons. The factorization proof for the DVCS should apply equally to the case of producing a virtual photon $\gamma^*_f$ with high $q_T$ and low virtuality $Q'^2$ that decays into a pair of charged leptons. Even if $q_T \gg Q'$, there is no large logarithm of $q_T / Q'$ that spoils perturbation theory, contrary to the inclusive process~\cite{Berger:2001wr}, because such logarithms are associated with diagrams' collinear sensitivity, which require two collinear parton lines to connect the low mass virtual photon to the hard part, which is suppressed by one power of $Q'/q_T$ compared to the direct photon attachment. In contrast, the DVMP amplitude has large logarithms of $q_T / m_D$, due to the long-distance evolution of the collinear parton lines. Such logarithms are incorporated by the evolution equation associated with the factorization formula in Eqs.~\eqref{eq:m2factorize} and \eqref{eq:p2 factorize}.

For a virtual photon $\gamma^*_f$ with its virtuality $Q'$ of the same order as $q_T$ (but, sufficiently away from masses of heavy quarkonia), it should belong to the short-distance hard part, and the whole process becomes $e^- + h \to h' + 2e^- + e^+$. 
This is no longer a $2\to 3$ SDHEP-type process, but we can still relate it to the SDHEP type by considering the kinematic regime
where one of the final-state electrons has a high transverse momentum $q_T$, balanced by the other $e^+e^-$ pair, which also has a large invariant mass $Q' \sim q_T$. 

First of all, the $\gamma^*$-mediated channel at $n = 1$ is allowed, with the hard scattering $e^- + \gamma^* \to 2e^- + e^+$. Second, the $n = 2$ channel does not unambiguously lead to the double DVCS (DDVCS) process~\cite{Guidal:2002kt} because it is not possible to distinguish which of the final-state electrons comes from the scattering of the initial-state electron.  
By labeling the final-state electrons and positron as $\pp{e^-_1, e^-_2, e^+}$, we find that a single configuration of $\pp{e^-_1, e^-_2, e^+}$ could correspond to both high-$Q'$ and low-$Q'$ processes. Specifically, let us consider the following three kinematic cases:
\begin{enumerate}
\item[(1)]
	All the $(e^-_1, e^-_2, e^+)$ have high transverse momenta, of order $q_T \gg \rt$, and the two invariant masses $( m_{e^-_1 e^+}, m_{e^-_2 e^+})$ are large, of the same order of $q_T$. This case leads unambiguously to DDVCS, and the factorization of DVCS can be trivially generalized here. But one needs to consider both diagrams with either $e^-_1$ or $e^-_2$ coming from the decay of the virtual photon $\gamma^*_f$.
\item[(2)]
	All the $(e^-_1, e^-_2, e^+)$ have high transverse momenta, of order $q_T \gg \rt$, but one of the invariant lepton-pair masses, say $m_{e^-_1 e^+}$, is much less than $q_T$, and the other pair has a large invariant mass, \ie, $q_T \sim m_{e^-_2 e^+} \gg m_{e^-_1 e^+}$. In this case, one can have (a) $(e^-_1, e^+)$ comes from the decay of a low-virtuality $\gamma^*_f$, and (b) $(e^-_2, e^+)$ comes from the decay of a high-virtuality $\gamma^*_f$. While both correspond to the DDVCS processes, it is the case (a) with a low-mass electron pair that contributes at a leading power.
\item[(3)]
	$(e_2^-, e^+)$ have high transverse momenta, of order $q_T \gg \rt$, and $e_1^-$ has a low transverse momentum, much less than $q_T$. Automatically, we have both $(m_{e^-_1 e^+}, m_{e^-_2 e^+})$ to be large. This gives two different cases: 
(a) $e_1^-$ comes from the diffraction of the initial-state electron, which gives out a quasireal photon $\gamma^*_{ee}$ that scatters with the diffractive hadron $h$ and produces a highly virtual photon $\gamma^*_f$ that decays into the $(e_2^-, e^+)$ pair; 
(b) $e_2^-$ comes from the hard scattering of the initial-state electron, whose interaction with the diffractive hadron $h$ produces a highly virtual photon $\gamma^*_f$ with a high transverse momentum, which decays into the $(e_1^-, e^+)$ pair.
Now only the case (b) corresponds to the DDVCS process, and case (a) gives a subprocess of (quasi)real photon scattering with the hadron, whose factorization will be proved later in Sec.~\ref{sec:a ll}. While both subprocesses are factorizable, it is the subprocess (a) that gives the leading power contribution.
\end{enumerate}
Of course, if the virtual photon $\gamma_f^*$ decays into a lepton pair of other flavors, like a $\mu^+\mu^-$ pair, then it unambiguously leads to the DDVCS process and can be factorized in the same way as the DVCS.

When the $\gamma_f^*$ virtuality $Q'$ becomes much greater than $q_T$, one starts entering the two-scale regime. Whether there will be large logarithms of $Q'/q_T$ that requires a new factorization theorem to be developed is not a trivial problem based on our analysis so far. We leave that discussion to the future.

For a heavy quarkonium production, unfortunately, it is not obvious that the factorization in Sec.~\ref{sec:e M} can be easily generalized. The key points to the factorization are
\begin{enumerate}
\item[(i)]
there is a pinch singularity that forces a collinear momentum to have the scaling in \eq{eq:col scaling}, with a leading component and two smaller components;
\item[(ii)]
soft gluons can be factorized from the collinear lines.
\end{enumerate}
The exclusive production of a heavy quarkonium naturally has the most contribution from producing a heavy quark pair with an invariant mass $M_H\sim 2m_Q$, where $m_Q \gg \LQCD$ is the heavy quark mass. Since the corresponding heavy quark GPD in $h$-$h'$ transition is suppressed, we do not suffer from the extra region like \fig{fig:p2 leading}(b). When the transverse momentum $q_T$ of the heavy quarkonium is much greater than $m_Q$, the heavy quark can be thought of as the active parton line associated with the observed particle $D$ in \fig{fig:p2 leading}(a), and the heavy quarkonium is attached to the hard part by a pair of nearly collinear heavy quark lines, whose momenta scale as
\beq
	k_Q \sim \pp{\lambda_Q ^2, 1, \lambda_Q } q_T,
	\quad
	\mbox{with } \lambda_Q = m_Q / q_T,
\eeq
when the heavy quarkonium moves along the minus direction. This pinches the plus momentum components to be small, and for a soft gluon $k_s$ attached to such heavy quark lines, one may keep only the $k_s^+$ component, which allows us to factorize the soft gluon out of the collinear lines. Hence, for $q_T \gg m_Q \gg \LQCD$, one can still factorize the heavy quarkonium production amplitude into the heavy quarkonium DA, up to the error of $\order{m_Q / q_T}$. See \paper{Kang:2014tta} for a similar discussion of the inclusive production of a heavy quarkonium.

When $m_Q \sim q_T\gg \LQCD$, the error estimated above becomes ${\cal O}(1)$, which invalidates the factorization into heavy quarkonium DA. 
However, if $M_H/2-m_Q\ll m_Q\sim q_T$, the formation of the heavy quarkonium from the produced heavy quark pair might be treated in terms of the color singlet model~\cite{Einhorn:1975ua, Chang:1979nn, Berger:1980ni} or the velocity expansion of nonrelativistic QCD with color singlet long-distance matrix elements~\cite{Bodwin:1994jh}.  For this exclusive production, the soft gluon interaction from the diffractive hadron with the heavy quark pair at $q_T\sim m_Q\gg \LQCD$ is expected to be suppressed by powers of $m_Qv/q_T\sim v$ with $v$ being the heavy quark velocity in the quarkonium's rest frame.  More detailed study for the heavy quarkonium production when $q_T \lesssim m_Q$ will be presented in a future publication. 

\subsection{Sensitivity to the $x$-dependence of GPDs}
\label{eq:e x}

As explained in the Introduction, the $x$-dependence of the GPD is very important while it is generally hard to extract for the following three reasons:
\begin{itemize}
\item[(1)] The factorization formalism is at amplitude level, and $x$ is the parton loop momentum, whose integration is always from $-1$ to $1$, and is never pinned down to a particular value.  For example, it is the integral
\beq[eq:dvcs hard]
	I_{\rm DVCS} = \int_{-1}^1 \dd{x} \frac{F(x,\xi,t)}{x \pm \xi \mp i\varepsilon}
\eeq 
to be evaluated when we compute the tree-level contribution to the DVCS scattering amplitude.

\item[(2)] The single-photon-mediated $n=1$ channel is generally allowed and its contribution could be more important than 
the GPD-sensitive $n=2$ subprocesses since the ratio of $n=1$ to $n=2$ channels is $Q/\rt$ power enhanced, even though the former might be suppressed by more powers of QED coupling constant. In addition, these two channels interfere with each other to make the extraction of GPDs harder.

\item[(3)] For most of the GPD-related processes, the hard coefficients only depend on $x$ and $\xi$ independently from the external observables, such that their convolution with the GPD only gives ``moment-type" information like \eq{eq:dvcs hard}.
\end{itemize}
The last two situations could be improved by identifying scattering processes that are more sensitive to the $x$-dependence of GPDs, whereas nothing can be done to alter the first dilemma. 

For the DVCS or DDVCS, the $q_T$ flow between the observed electron and final-state photon $\gamma(q_2)$ (or virtual photon $\gamma_f^*(q_2)$ in the case of DDVCS) goes through the $\gamma_{ee}^*(q)$ and the quark line of momentum $\widetilde{q}$ (in red) in Fig.~\ref{fig:dvcs LO} at the leading order. From the invariant mass of the final-state photon (or the virtual photon),
\beq[eq:q2sq]
	q_2^2 = \left( 2\xi P+q \right)^2 = (2\xi) 2P\cdot q-Q^2+{\cal O}(|t|)\, ,
\eeq
with terms of ${\cal O}(|t|/Q^2)$ neglected, we can express the invariant mass of the exchanged quark line of momentum $\widetilde{q}$ in Fig.~\ref{fig:dvcs LO} as
\begin{eqnarray}
	\widetilde{q}^{\, 2} &=& 
		\left( (x+\xi)P +q \right)^2 
			\nonumber\\
	&=& \frac{Q^2+q_2^2}{2\xi}
	\left[x- \xi \left(\frac{1-q_2^2/Q^2}{1+q_2^2/Q^2}\right)\right],
\label{eq:qqsq}
\end{eqnarray}
which is proportional to $x - \xi$ when $q_2^2\to 0$, leading to the well-known $1/(x-\xi+i\varepsilon)$ structure for the leading-order hard coefficient of DVCS. On the other hand, the leading-order hard coefficient for DDVCS with $q_2^2 \neq 0$ provides a direct link between the loop momentum fraction $x$ and the invariant mass of the observed lepton pair $q_2^2$, as shown in Eq.~(\ref{eq:qqsq}).
It is this connection between $x$ and the externally measured observable $q_2^2$ that provides the enhanced sensitivity to the $x$-dependence of GPDs beyond the moment type~\cite{Guidal:2002kt}.

For the DVMP, the $q_T$ flow between the final-state meson $M_D$ and scattered electron goes through the $\gamma_{ee}^*$ as well as the parton lines, as shown in Fig.~\ref{fig:dvmp lo}(b). The hard coefficient is similar to that of the pion electromagnetic form factor and only provides a moment-type sensitivity.

In addition, the $n=1$ virtual photon $\gamma^*$-mediated subprocesses are allowed for DVCS and DDVCS, and they are at the same perturbative order as the leading $n=2$ GPD-induced channels. For DVMP, only the production of charge-neutral $C$-even mesons allows the $\gamma^*$-mediated $n=1$ channel, which is of higher order in QED coupling constant comparing to those GPD-sensitive $n=2$ channels. The $\gamma^*$-mediated channels at the amplitude level interfere with the GPD-sensitive channels and need to be carefully treated for extracting the $x$-dependence of GPDs.

\section{SDHEP with a real photon beam}
\label{sec:photon}

For single diffractive hard exclusive photoproduction processes, we have $B = \gamma$. The other particles $C$ and $D$ can be two elementary particles, one elementary particle and one light meson, or two light mesons. 
So we consider the three cases: 
(1) massive dilepton $(CD) = (l^+l^-)$~\cite{Berger:2001xd, CLAS:2021lky} or diphoton $(\gamma\gamma)$ production~\cite{Pedrak:2017cpp, Grocholski:2021man, Grocholski:2022rqj}, 
(2) real photon and light meson pair $(CD) = (\gamma M_D)$ production~\cite{Boussarie:2016qop, Duplancic:2018bum},
and 
(3) light meson pair $(CD) = (M_C M_D)$ production~\cite{ElBeiyad:2010pji}. 
In this section, we provide the factorization arguments for all these processes by following the same strategy used for factorization of the SDHEP with a lepton beam in terms of the two-stage paradigm, presented in Eqs.~\eqref{eq:diffractive}--\eqref{eq:channels}.

\subsection{Massive dilepton or diphoton production: $(CD) = (l^+l^-)$ or $(\gamma\gamma)$}
\label{sec:a ll}

Both production processes allow the $\gamma^*$-mediated $n=1$ subprocesses. For the dilepton production, we have the partonic process $\gamma\gamma^* \to l^+ l^-$, starting at $\order{e^2}$ in terms of the QED coupling $e$, while we have $\gamma\gamma^* \to \gamma\gamma$ for the diphoton production, starting at $\order{e^4}$. Since this $\gamma^*$-mediated $n=1$ channel has a power enhancement of $\order{q_T/\rt}$ compared to the $n = 2$ channel, it cannot be simply neglected even though its scattering amplitude might require a higher power in QED coupling. A careful quantitive comparison in size between $\gamma^*$-mediated $n=1$ and GPD-sensitive $n=2$ subprocesses is needed in practical evaluation.

\begin{figure}[htbp]
\centering
	\begin{tabular}{cc}
		\includegraphics[scale=0.7]{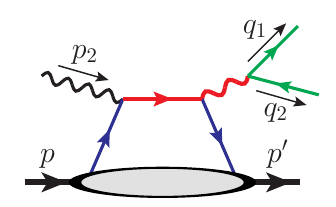} &
		\includegraphics[scale=0.7]{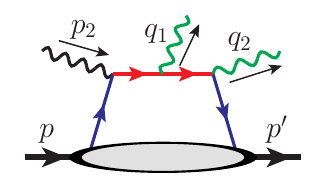}	\\
		(a) & (b) 
	\end{tabular}
\caption{Examples of leading-order diagrams in the GPD channel for the single diffractive hard exclusive photoproduction of massive (a) dilepton and (b) diphoton processes.}
\label{fig:a2ll a2aa}
\end{figure}

For $n = 2$ channel, these two processes share the same feature as the DVCS, as well as the same leading-region graphs in \fig{fig:dvcs} with a proper change of the external lines, because $B$, $C$, and $D$ are all elementary colorless particles.
The argument for factorization into GPDs works in the same way as for the DVCS in Sec.~\ref{sec:dvcs} and will not be repeated here. The process with $(CD) = (l^+l^-)$ happens by producing a timelike photon $\gamma^{\prime*}$ in the exclusive $\gamma h \to \gamma^{\prime*} h'$ process followed by the decay $\gamma^{\prime*} \to l^+l^-$, which is the TCS process, as shown in \fig{fig:a2ll a2aa}(a). For the process with $(CD) = (\gamma\gamma)$, all the three photons couple to the quark lines, as illustrated in \fig{fig:a2ll a2aa}(b).
In both processes, it is the high $q_T$ that provides the hard scale for factorizability, by creating high virtualities through the invariant mass of the virtual photon in the dilepton case or having the $q_T$ flow through the quark lines in the diphoton case.

\begin{figure}[htbp]
\centering
	\begin{tabular}{cc}
		\includegraphics[scale=0.7]{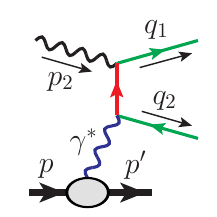} &
		\includegraphics[scale=0.7]{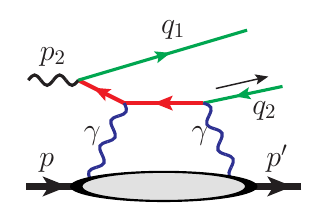}	\\
		(a) & (b) 
	\end{tabular}
\caption{(a) The sample diagram for the $\gamma^*$-mediated channel of the photoproduction of a massive lepton pair, where the internal lepton propagator (in red) has a hard virtuality only when $q_T$ is large. (b) At large $m_{ll}$ but small $q_T$, the forward scattering diagrams with two photon exchanges between the diffractive hadron and the quasireal lepton can become important and compete with the TCS mechanism in \fig{fig:a2ll a2aa}(a).}
\label{fig:a2ll a}
\end{figure}

It is important to note that in general, the requirement of a high invariant mass for the pair of particles $(CD)$ is not the same as requiring a hard $q_T$. For the TCS, it is the invariant mass of the lepton pair $m_{ll}$ that provides the hard scale for the partonic collision, and hence keeping $m_{ll}$ large is sufficient for TCS to be factorized into GPD, independent of the magnitude of $q_T$ of the observed lepton. However, a hard $q_T$ is needed to guarantee the $\gamma^*$-mediated $n = 1$ subprocess $\gamma\gamma^* \to l^+ l^-$ to be a hard scattering process, as illustrated in \fig{fig:a2ll a}(a). If $q_T$ is too low, then this amplitude introduces another enhancement factor of $\order{m_{ll} / q_T}$, in addition to the $m_{ll} / \rt$ enhancement of the $n = 1$ channel, as correctly pointed out in \paper{Berger:2001xd}. Then, this could allow other subprocesses to happen that may compete with the TCS subprocess in magnitude. For example, one may have an $n = 2$ channel mediated by $f_2 = [\gamma\gamma]$, as shown in \fig{fig:a2ll a}(b), which is suppressed by $e^2$ and one power of $\rt / m_{ll}$ compared to the $n = 1$ channel, but is still one power $\order{m_{ll} / q_T}$ higher than the TCS channel. The relative order comparison is then too complicated to be obvious, and the extraction of GPDs from the TCS amplitude becomes even harder. 

\begin{figure}[htbp]
\centering
	\includegraphics[scale=0.7]{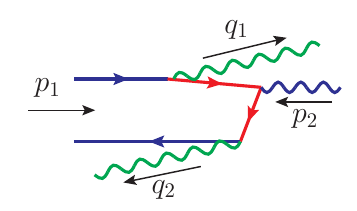}
\caption{A sample diagram for the photoproduction of diphoton process at low $q_T$, where the photon $q_1$ is radiated collinearly by the incoming quark.}
\label{fig:a2aa col}
\end{figure}

On the other hand, if $q_T$ is too low in the diphoton production process, some quark lines could have low virtualities of order $q_T$, as the photons could be radiated from the quark lines (see \fig{fig:a2aa col}) almost collinearly, introducing the long-distance physics into the ``hard probe'', which invalidates our factorization arguments.

\subsection{Real photon and light meson pair production: $(CD) = (\gamma M_D)$}
\label{sec: a aM}
For $(CD) = (\gamma M_D)$ with $M_D$ being a light meson, the $n = 1$ channel corresponds to the subprocess $\gamma^* \gamma \to \gamma M_D$. This is forbidden for a charged meson like $\pi^{\pm}$, as considered in \paper{Duplancic:2018bum}, or for a neutral meson with even $C$-parity, like $\pi^0$, $\eta$, etc. In the high-$q_T$ scattering, the $n = 1$ amplitude can be factorized into the DA of $M_D$.  

The $n = 2$ channel has the same color structure as the DVMP process in Sec.~\ref{sec:e M}, and the leading region is also as in \fig{fig:p2 leading} just with the proper change of the external electron lines by photon lines. The argument for factorization then works in the same way as for the DVMP, and is not to be repeated here. For the same reason as the diphoton production process in the previous subsection, we emphasize the necessity of the hard transverse momentum $q_T$, which is not equivalent to requiring a large invariant mass of the $\gamma M_D$ pair.

\subsection{Light meson pair production: $(CD) = (M_C M_D)$}
\label{sec:a MM}
The single diffractive photoproduction with $(CD) = (M_C M_D)$ differs from the electroproduction of a light meson in Sec.~\ref{sec:e M} by having one more hadron in the final state. This leads to one more collinear subgraph in another direction but does not make the factorization proof very different. By focusing on the difference, we will present the factorization proof as a generalization of that for the DVMP in Sec.~\ref{sec:e M}, and especially, we will make essential use of the asymmetric contour deformation as explained there.

The $n = 1$ channel is given by the subprocess $\gamma^* \gamma \to M_C M_D$, which may or may not happen depending on the quantum numbers of $M_C$ and $M_D$. This was considered first in \paper{Brodsky:1981rp} and the time-reversed process was also studied in \paper{Qiu:2022bpq}. The amplitude can be factorized into the DAs of $M_C$ and $M_D$, whose discussion we refer to \paper{Brodsky:1981rp}.

\begin{figure}[htbp]
\centering
	\begin{tabular}{cc}
		\includegraphics[scale=0.6]{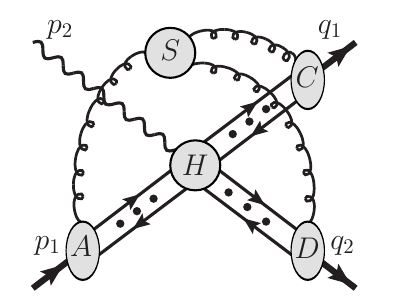} &
		\includegraphics[scale=0.6]{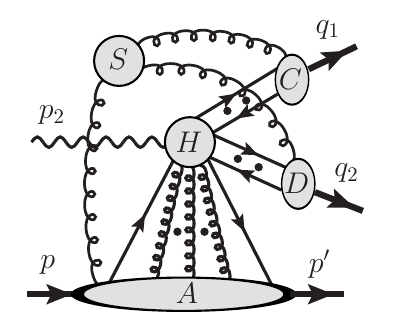} 	\\
		(a) & (b) 
	\end{tabular}
\caption{Leading-region graphs for the photoproduction of a light meson pair. (a) is for the $2\to2$ hard exclusive scattering of SDHEP, and (b) is for the full SDHEP. There can be any numbers of soft gluons connecting $S$ to each collinear subgraph. The regions with $S$ connecting to one or more collinear subgraphs via quark lines or transversely polarized gluon lines are omitted. Depending on the quantum numbers, the collinear quark lines may be replaced by transversely polarized gluon lines. The dots represent arbitrary numbers of longitudinally polarized collinear gluons.}
\label{fig:a MM leading}
\end{figure}

For the $n = 2$ channel, the leading regions are shown in \fig{fig:a MM leading} for the $2\to2$ hard exclusive scattering of the SDHEP in (a) and the full SDHEP in (b), where we assumed that all lines in the hard part ``$H$'' are off shell by order of the hard scale $Q\sim q_T$, which effectively makes the contribution from attaching soft gluons to $H$ power suppressed. There could be additional leading regions in which one or more of the collinear subgraph is connected to the soft subgraph by one quark or transversely polarized gluon line, while connecting to the hard subgraph by the other quark or transversely polarized gluon line. Following the same assumption that such soft end point region is strongly suppressed by the nonperturbative QCD dynamics from the meson distribution amplitude, we neglect them and consider only the leading regions in \fig{fig:a MM leading}.

\subsubsection{The $2\to2$ hard exclusive scattering of SDHEP}
\label{sec:a MM2}
First, we examine the $2\to2$ hard exclusive scattering of the SDHEP with a photon beam,
\beq[eq:m3 2to2]
	M_A(p_1) + \gamma(p_2) \to M_C(q_1) + M_D(q_2),
\eeq
in the c.m. frame. By generalizing \eq{eq:dvmp aux vectors}, we introduce three sets of light-cone vectors,
\begin{align}
	&w_A^{\mu} = \frac{1}{\sqrt{2}}(1, \hat{z}),
	&&\bar{w}_A^{\mu} = \frac{1}{\sqrt{2}}(1, -\hat{z}),	\nn\\
	&w_C^{\mu} = \frac{1}{\sqrt{2}}(1, \hat{n}) = \bar{w}_D^{\mu},
	&&\bar{w}_C^{\mu} = \frac{1}{\sqrt{2}}(1, -\hat{n}) = w_D^{\mu},
\end{align}
where $\hat{z}$ and $\hat{n}$ are normalized three-vectors along the directions of the initial-state meson $M_A$ and final-state meson $M_C$. Basically, $w_{A,C,D}$ are the light-cone vectors along the directions of meson $A$, $C$ and $D$, respectively, and the corresponding vectors with bars refer to the conjugate light-cone vectors along the opposite directions. The essential point is that any soft gluon momentum $k_s$ can be routed to only flow through two collinear subgraphs. For this, we introduce the notation $k_s^{(ij)}$ to be a soft gluon momentum that leaves the collinear subgraph $i$ into $S$, then into the collinear subgraph $j$, through the hard subgraph $H$ and back to $i$. Apparently, we have $k_s^{(ij)} = -k_s^{(ji)}$, with $i, j = A, C, D$ and $i \neq j$.

When considering the soft gluon momentum $k_s^{(ij)}$, we expand it in the $w_i$-$w_j$ frame as defined in \eq{eq:frame},\footnote{While we may define the plus and minus components in each $w_i$-$w_j$ frame like Eqs.~\eqref{eq:frame}-\eqref{eq:cov scaling}, having multiple such frames makes the notation cumbersome, so we stick to the covariant notations.}
\beq
	k_s^{(ij)} = w_i \frac{k_s^{(ij)}\cdot w_j}{w_i \cdot w_j} + w_j \frac{k_s^{(ij)}\cdot w_i}{w_i \cdot w_j} + k_{sT}^{(ij)} \,,
\eeq
where all the three terms on the right are of the same size, $\order{\lambda_s Q}$.
When it flows in the collinear subgraph $i$, whose momenta are dominantly along $w_i$, the $k_s^{(ij)}$ can be approximated to only retain the $w_j$ component, 
\beq[eq:soft kij]
	k_s^{(ij)} \simeq \hat{k}_s^{(ij)} = w_j \frac{k_s^{(ij)}\cdot w_i}{w_i\cdot w_j}	\,.
\eeq
Moreover, the coupling of this soft gluon to the collinear subgraph $\D^i$ can be approximated as
\begin{align}\label{eq:soft coupling ij}
	&\D^i_{\mu}(k_i, k_s^{(ij)}) \, g^{\mu\nu} \, S_{\nu}(k_s^{(ij)})		\nn\\
		&\hspace{3em}
		\simeq 
		\D^i_{\mu}(k_i, \hat{k}_s^{(ij)}) \, \frac{\hat{k}_s^{(ij)\mu} \, w_i^{\nu}}{k_s^{(ij)} \cdot w_i} \, S_{\nu}(k_s^{(ij)})	\,,
\end{align}
because it is the component $g^{-+}$ of $g^{\mu\nu}$, which is given by $w_j^{\mu} w_i^{\nu} / w_i \cdot w_j$, that provides the dominant contribution. 
In \eq{eq:soft coupling ij}, $k_i$ stands for some collinear momentum in the subgraph $i$.
This approximation will allow the use of Ward identity to factorize the soft gluons out of the collinear subgraphs.\footnote{We should note that the argument given here is equivalent to Refs.~\cite{Collins:1981ta, Nayak:2005rt} that boost into the rest frame of two collinear subgraphs. The underlying reason is that any two distinct collinear subgraphs are well separated in rapidity; in the language here, it is $w_i \cdot w_j \simeq \order{1}$.}

While this is a good approximation for the central soft region, it is not for the Glauber region in which 
\beq[eq:glauber ij]
	| k_s^{(ij)}\cdot w_i | \, | k_s^{(ij)}\cdot w_j | \ll | k_{sT}^{(ij)} |^2 \, w_i \cdot w_j \,.
\eeq
Now because all the collinear lines in the subgraph $i$ or $j$ only give poles for $k_s^{(ij)}\cdot w_i$ or $k_s^{(ij)}\cdot w_j$
on the same half complex plane, the integration contour of $k_s^{(ij)}$ is not pinched in the Glauber region, and a proper deformation can get it out of the Glauber region.
Since we anticipate that in generalizing the factorization to the diffractive process in the next subsection, a soft momentum $k_s^{(Aj)}$ flowing in the $A$-collinear subgraph has its component $k_s^{(Aj)} \cdot w_A$ trapped in the Glauber region, we choose not to deform the contour of $k_s^{(Aj)} \cdot w_A$ in this $2\to2$ hard exclusive scattering of the full SDHEP while trying to factorize soft interactions from other leading collinear subgraphs.

The needed deformations can be motivated by examining a single soft gluon exchange between different collinear subgraphs. We first consider the collinear subgraph $C$ that has one soft gluon $k_s^{(CA)}$ and $k_s^{(CD)}$ exchange with the $A$-collinear subgraph and $D$-collinear subgraph, respectively. Since $k_s^{(CA)}$ flows in $C$ in the same direction as the $C$-collinear lines, the poles of $k_s^{(CA)} \cdot w_C$ are all on the lower half complex plane, so we deform the contour of $k_s^{(CA)}$ by
\beq[eq:deform CA]
	k_s^{(CA)} \to k_s^{(CA)} + i \, w_A \, \order{Q}  \,,
\eeq
when it is in the Glauber region. Similarly, we deform the contour of $k_s^{(CD)}$ by
\beq[eq:deform CD]
	k_s^{(CD)} \to k_s^{(CD)} + i \, w_D \, \order{Q}  \,.
\eeq
In order for the approximator in \eq{eq:soft coupling ij} not to obstruct such deformations, we modify it to
\bse\label{eq:soft coupling C}
\begin{align}
	&\D^C_{\mu}(k_C, k_s^{(CA)}) \, g^{\mu\nu} \, S_{\nu}(k_s^{(CA)})		\nn\\
		&\hspace{1em}
		\simeq 
		\D^C_{\mu}(k_C, \hat{k}_s^{(CA)}) \, 
			\frac{\hat{k}_s^{(CA)\mu} \, w_C^{\nu}}{k_s^{(CA)} \cdot w_C + i\varepsilon} \, 
			S_{\nu}(k_s^{(CA)})	\,,	\label{eq:soft coupling CA}	\\
	&\D^C_{\mu}(k_C, k_s^{(CD)}) \, g^{\mu\nu} \, S_{\nu}(k_s^{(CD)})		\nn\\
		&\hspace{1em}
		\simeq 
		\D^C_{\mu}(k_C, \hat{k}_s^{(CD)}) \, 
			\frac{\hat{k}_s^{(CD)\mu} \, w_C^{\nu}}{k_s^{(CD)} \cdot w_C + i\varepsilon} \, 
			S_{\nu}(k_s^{(CD)})	\,, \label{eq:soft coupling CD}
\end{align}
\ese
where only the relevant arguments are written explicitly.
Both approximations in \eq{eq:soft coupling C} have the structure
\begin{align}
	&\D^C_{\mu}(k_C, k_s) \, g^{\mu\nu} \, S_{\nu}(k_s)		\nn\\
		&\hspace{3em}
		\simeq 
		\D^C_{\mu}(k_C, \hat{k}_s) \, 
		\frac{\hat{k}_s^{\mu} \, w_C^{\nu}}{k_s \cdot w_C + i\varepsilon} \, 
		S_{\nu}(k_s)	\,,
\end{align}
where the structure $\hat{k}_s^{\mu} \, \D^C_{\mu}(k_C, \hat{k}_s)$ allows the use of Ward identity in a uniform way, no matter which other collinear subgraph $k_s$ flows through. The $+ i\varepsilon$ choice will lead to future-pointing soft Wilson lines.

Now we consider the collinear longitudinally polarized gluons attaching $C$ to $H$. Similarly, the approximation can be obtained by examining a single gluon, whose momentum $k_C$ flows from $H$ into $C$ and can be expanded in the $w_C$-$\bar{w}_C$ frame,
\beq
	k_C = w_C \, (k_C \cdot \bar{w}_C) + \bar{w}_C \, (k_C \cdot w_C) + k_{C,T} \, ,
\eeq
where among the three terms on the right, the $w_C$ component dominates and scales as $\order{Q}$. Then we approximate $k_C$ in $H$ by
\beq
	k_C \to  \hat{k}_C = w_C \, (k_C \cdot \bar{w}_C) \,,
\eeq
and the coupling of the collinear gluon to $H$ by
\begin{align}\label{eq:collinear approx C}
	&H_{\mu}(k_H, k_C) \, g^{\mu\nu} \, \D^C_{\nu}(k_C)		\nn\\
		&\hspace{3em}
		\simeq 
		H_{\mu}(k_H, \hat{k}_C) \,
		\frac{\hat{k}_C^{\mu} \, \bar{w}_C^{\nu} }{k_C \cdot \bar{w}_C - i\varepsilon} \,
		\D^C_{\nu}(k_C)	\,,
\end{align}
where only the relevant argument dependence is written explicitly and $k_H$ stands for some hard momentum in $H$. 

The $- i\varepsilon$ in Eq.~(\ref{eq:collinear approx C}) is chosen in order to be compatible with the deformations in Eqs.~\eqref{eq:deform CA} and \eqref{eq:deform CD}. Even though we are approximating the collinear region, which does not suffer from the Glauber region problem, \eq{eq:collinear approx C} is applied to the whole diagram with deformed contours. Furthermore, the same gluon $k_C$ considered in \eq{eq:collinear approx C} can also go into the soft region, attaching to $A$- or $D$-collinear subgraph, for which we will change the notation $k_C$ to $k_C^{(A)}$ or $k_C^{(D)}$,\footnote{Note that now the soft momentum direction is reversed compared to the convention of $k_s^{(CA)}$ and $k_s^{(CD)}$, which are used in Eqs.~\eqref{eq:deform CA} and \eqref{eq:deform CD}.} 
whose contribution has already been included in the soft approximations defined in \eq{eq:soft coupling C}. A subtraction is needed from \eq{eq:collinear approx C} to avoid such double counting, which is obtained by first applying the soft approximation [\eq{eq:soft coupling C}] and then applying the collinear approximation [\eq{eq:collinear approx C}]. Since the subtraction mixes the collinear and soft approximations for the same gluons, and the latter require deformation of contours, we do need the $i\varepsilon$ prescription in \eq{eq:collinear approx C} not to obstruct the contour deformations in Eqs.~\eqref{eq:deform CA} and \eqref{eq:deform CD}. Since we need the deformations
\beq
	\Delta k_C^{(A)} = -i \, w_A \, \order{Q},
	\quad
	\Delta k_C^{(D)} = -i \, w_D \, \order{Q},
\eeq
which means that the denominator in \eq{eq:collinear approx C} needs to be compatible with the deformations
\begin{align}
	\Delta k_C^{(A)} \cdot \bar{w}_C &= -i \, (w_A\cdot \bar{w}_C) \, \order{Q} = -i  \, \order{Q}	\,,	\nn\\
	\Delta k_C^{(D)} \cdot \bar{w}_C &= -i \, (w_D\cdot \bar{w}_C) \, \order{Q} = 0	\,.
\end{align}
This explains the $- i\varepsilon$ choice in \eq{eq:collinear approx C}. After applying Ward identity, it leads to collinear Wilson lines pointing to the past.

Equations~\eqref{eq:soft coupling C} and \eqref{eq:collinear approx C} constitute the needed approximations related to the collinear subgraph $C$.
Even though we only considered a single soft or collinear gluon connection, they generalize to multiple gluon connections in an obvious way: one just applies \eq{eq:soft coupling C} to every soft gluon connecting $C$ to $A$ or $D$, and \eqref{eq:collinear approx C} to every collinear longitudinally polarized gluon connecting $H$ to $C$. Then by applying suitable on-shell projections to the $C$-collinear quark lines or transversely polarized gluon lines, and summing over all possible attachments of the collinear gluons, we can factorize the collinear longitudinally polarized gluons out of the hard part $H$ onto two Wilson lines along $\bar{w}_C$ pointing to the past, and the soft gluons out of $C$ onto two Wilson lines along $w_C$ pointing to the future.

We should note that by choosing lightlike auxiliary vectors $w_C$ in the soft approximation \eq{eq:soft coupling C}, the resultant soft factor contains rapidity divergences. This can be remedied by a different vector $n_C$ that differs from $w_C$ by being slightly off light cone, as in Refs.~\cite{Collins:2011zzd, Qiu:2022bpq} for example, which does not affect the argument. Since the soft gluons eventually cancel whether we use $w_C$ or $n_C$, the problem of rapidity divergence does not affect our argument of collinear factorization, and we will simply use the lightlike vector $w_C$.

\begin{figure}[htbp]
\centering
	\begin{tabular}{cc}
		\includegraphics[scale=0.6]{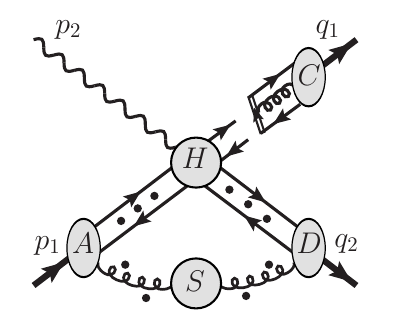} &
		\includegraphics[scale=0.6]{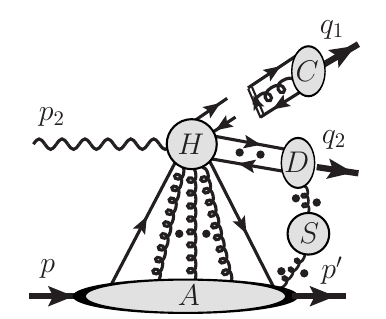} 	\\
		(a) & (b) 
	\end{tabular}
\caption{The factorization of the $C$-collinear subgraph out of the hard subgraph $H$ into the $C$ meson DA, and soft gluons out of the $C$-collinear subgraph, which results in their cancellation. The soft subgraph is now only coupled to the $D$-collinear subgraph and (a) the effective ``meson'' state $A$, or (b) the single diffractive hadron $h\to h'$.}
\label{fig:a MM factorize C}
\end{figure}

The subsequent argument follows the same line of \paper{Qiu:2022bpq}. The essence is that the factorized soft Wilson lines along $w_C$ are only coupled to the factorized collinear subgraph $C$ in colors, but not in momenta and Lorentz indices. The exclusiveness guarantees that the collinear factor $\D^C$ is a color singlet and becomes a meson DA, which then ensures the cancellation of the soft gluons coupled to $C$. This reduces the graph in \fig{fig:a MM leading}(a) to the partly factorized one in \fig{fig:a MM factorize C}(a), in which only the two collinear subgraphs $A$ and $D$ are coupled to the hard subgraph $H$, and the soft subgraph $S$ is only coupled to $A$ and $D$ subgraphs.

With the $C$-collinear subgraph factorized out, the leading-region graph in \fig{fig:a MM factorize C}(a) is similar to that in \fig{fig:m2 leading}(a), whose factorization is shown in Sec.~\ref{sec:DVMP m}. Again, in the treatment of the soft region, one only needs to deform the contour of soft gluon $k_s^{(DA)}$ by 
\beq
	k_s^{(DA)} \to k_s^{(DA)} + i \, w_A \, \order{Q},
\eeq
regardless of the poles of $k_s^{(DA)} \cdot w_A$ provided by the $A$-collinear propagators. By the same argument as for the $C$ subgraph, the soft gluons coupling to $D$ are canceled, and the $D$ subgraph is factorized out of $H$ into the meson DA for $D$. Then the soft gluons are only coupled to the $A$ subgraph and no longer pinched. They can then be deformed into the $A$-collinear region and grouped into a part of $A$-collinear subgraph, which can be further factorized from $H$ into the DA of the $A$ meson. 

Finally, the amplitude of the $2\to 2$ scattering in \eq{eq:m3 2to2} is factorized as
\begin{align}\label{eq:m3factorize}
	&\M_{M_A \gamma \to M_C M_D} = \sum_{i, j, k} \int_0^1 \dd{z_A} \dd{z_C} \dd{z_D} \phi_{i/A}(z_A)  \nn\\
		& \hspace{1em} \times C_{i \gamma \to j k} (z_A, z_C, z_D; q_T) \, \phi_{j/C}(z_C) \, \phi_{k/D}(z_D),
\end{align}
where $\phi_{i/A}(z_A)$, $\phi_{j/C}(z_C)$, and $\phi_{k/D}(z_D)$ are the DAs associated with the initial-state meson $M_A$ and final-state mesons $M_C$ and $M_D$, respectively, the $i, j$, and $k$ run over all possible parton channels, $[q\bar{q}']$ or $[gg]$, as well as their spin structures, and the hard coefficient $C_{i \gamma \to j k}$ is the scattering between a photon and parton pair $i$ into two parton pairs $j$ and $k$.

\subsubsection{Production of two high-$q_T$ light mesons in the SDHEP with a photon beam}
Extending the factorization of the previous subsection to the full SDHEP of two high-$q_T$ light mesons is trivial. The only complication arises from the extra DGLAP region in the single diffractive channel of the hadron $h\to h'$, which causes the momentum $k_s$ of the soft gluon coupling to the $A$-collinear subgraph to be pinched in the Glauber region for its component $k_s \cdot w_A$, as explained in Sec.~\ref{sec:DVMP full}. The strategy that we used in the previous subsection for factorizing the SDHEP's hard exclusive $2\to2$ scattering applies here with no change, because we never deformed the contour of $k_s \cdot w_A$ when $k_s$ flows through the $A$-collinear subgraph. The important step of factorizing the $C$-collinear subgraph is shown in \fig{fig:a MM factorize C}(b). In the end, the diffractive amplitude is factorized into the hadron GPD and meson DAs,
\begin{align}\label{eq:p3 factorize}
	&\mathcal{M}^{(2)}_{h\gamma \to h' M_C M_D} = \sum_{i, j, k} \int_{-1}^1 \dd{x} \int_0^1 \dd{z_C} \dd{z_D} F^{hh'}_i(x, \xi, t) 
	\nn\\
	&\hspace{1em}\times
	C_{i\gamma \to jk}(x, \xi; z_C, z_D; q_T) \, \phi_{j/C}(z_C) \, \phi_{k/D}(z_D) ,
\end{align}
up to $1/q_T$ power suppressed terms, where the symbols have the same definitions as those in \eq{eq:m3factorize}, and the hard coefficient $C_{i\gamma \to jk}(x, \xi; z_C, z_D; q_T)$ can be calculated in the same way as that in \eq{eq:m3factorize}, just with a proper variable change $z_A \to z_A(x, \xi) = (x + \xi)/(2\, \xi)$ as in \paper{Qiu:2022bpq}.

\subsection{Virtual photon or heavy quarkonium production}
\label{subsec:a virtual photon}
We have considered the production of real photons from the SDHEP with a photon beam, like the diphoton production or photon-meson pair production discussed above. At leading power, the produced photons in the final state directly couple to the hard interaction. We can also have any one (or both) of those photons being virtual with an invariant mass $Q'$, which can vary from $Q' \ll q_T$ to $Q' \lesssim q_T$. For the same reason as stated in Sec.~\ref{subsec:e virtual photon}, the leading-power diagrams also have those virtual photons directly coupling to the hard part. Therefore, factorization works in the same way as for the real photon cases as long as the virtual photons still have hard transverse momenta $q_T$. The scale $Q$ of the hard part now depends on both $q_T$ and the photon virtuality $Q'$, schematically as $Q \sim \sqrt{q_T^2 + Q'^2}$. 

The kinematic signal for the virtual photon is a charged lepton pair $l^+l^-$ with a hard total transverse momentum $q_T$. Depending on the relative size between $q_T$ and their invariant mass $m_{ll}$, the hard scattering amplitudes take different forms in a way similar to the analysis in Sec.~\ref{subsec:e virtual photon}. We leave a detailed study to the future.

While the factorization for real photon production can be directly generalized to virtual photon production, going from the light meson production to a heavy quarkonium production is not so trivial, as explained in Sec.~\ref{subsec:e virtual photon}. One situation where one can still factorize the heavy quarkonium production into its DA is when $q_T$ is much greater than the heavy quark mass $m_Q$, in which case the error of the factorization is enhanced from power of $\LQCD / q_T$ to power of $m_Q / q_T$. But the factorizability when $q_T$ is of the same order as $m_Q$ needs further study, as in Sec.~\ref{subsec:e virtual photon}.

\subsection{Sensitivity to the $x$-dependence of GPDs}
\label{eq:p x}
As discussed in Sec.~\ref{eq:e x}, to get sensitivity to the $x$-dependence of GPDs, we need to find SDHEPs
that avoid the $n = 1$ channel and have their partonic hard parts to include the $x$-dependence that cannot be factorized from the external observables. 

For both the dilepton or diphoton production processes, one has the $n = 1$ channel, and the hard coefficient of the $n = 2$ channel takes a factorized form~\cite{Berger:2001xd, Pedrak:2017cpp}, so that these processes can only provide moment-type sensitivity, which is further contaminated by the $\gamma^*$-mediated $n=1$ subprocesses.

\begin{figure}[htbp]
\centering
	\begin{tabular}{cc}
		\includegraphics[scale=0.6]{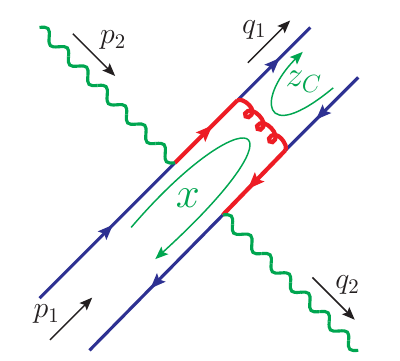} &
		\includegraphics[scale=0.6]{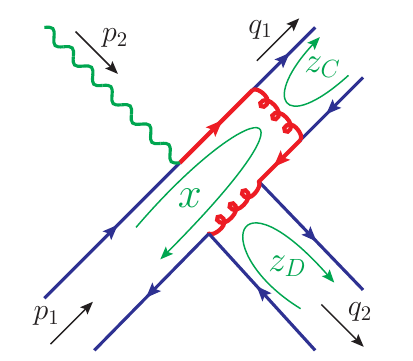} \\
		(a) & (b)
	\end{tabular}
	\caption{Example diagrams of the hard parts for (a) the single diffractive photoproduction of photon-meson pair, and (b) the single diffractive photoproduction of meson pair. The parton momentum flows are indicated by green thin curved arrowed lines, and the amputated parton lines are indicated in blue.}
\label{fig:a hard}
\end{figure}

For the $\gamma$-meson pair production, which is a crossing process of the meson-production of diphoton process considered in \paper{Qiu:2022bpq}, we show an example of the hard-part diagrams in \fig{fig:a hard}(a). Similar to the type-$A$ diagrams of the diphoton production process in \paper{Qiu:2022bpq}, due to the gluon propagator that connects the two fermion lines, the hard coefficients depend on both $x$ and $q_T$ in a nontrivial way to provide enhanced sensitivity to the $x$-dependence of GPDs. Also, by a suitable choice of the final-state meson species, one can eliminate the $n = 1$ channel; see the discussion in Sec.~\ref{sec: a aM}.

As for the meson pair production process, the leading-order $[q\bar{q}']$-channel diagrams of the hard scattering contain two quark and two gluon propagators, shown in \fig{fig:a hard}(b) for example. The gluon propagator with both $x$ and $z_C$ flowing through is exactly the same as that in \fig{fig:a hard}(a) for the photon-meson pair production process. For the same reason, it leads to enhanced sensitivity to the $x$-dependence of GPDs.

A more detailed discussion of the sensitivity to GPD $x$-dependence will be given later in Sec.~\ref{sec:x-dependence}, together with a general criterion for the enhanced sensitivity. 

\section{SDHEP with a meson beam}
\label{sec:meson}

For the SDHEP with a meson beam, we have $B$ being some meson $M_B$, which is usually a pion or kaon. Similar to the case with a photon beam, we consider three cases for the particles $C$ and $D$: (1) massive dilepton $(CD) = (l^+l^-)$ or diphoton $(\gamma\gamma)$ production; (2) real photon and light meson pair $(CD) = (\gamma M_D)$ production; and (3) light meson pair $(CD) = (M_C M_D)$ production. The dilepton and diphoton production processes have been studied in Refs.~\cite{Berger:2001zn, Qiu:2022bpq}, respectively, and their factorizations are similar to the DVMP process. The processes (2) and (3) have not been considered in the literature. In this section, we address the factorization of these processes in the framework of the SDHEP within the two-stage paradigm.

\subsection{Massive dilepton or diphoton production: $(CD) = (l^+l^-)$ or $(\gamma\gamma)$}

The SDHEPs of massive dilepton and diphoton productions are
\beq[eq:M ll]
	h(p) + M_B(p_2) \to h'(p') + l^-(q_1) + l^+(q_2),
\eeq
and 
\beq[eq:M aa]
	h(p) + M_B(p_2) \to h'(p') + \gamma(q_1) + \gamma(q_2),
\eeq
respectively. 
Both processes have $C$ and $D$ being colorless elementary particles, and they are similar to the meson production in the SDHEP with a lepton beam in Sec.~\ref{sec:e M} and the meson-photon pair production in the SDHEP with a photon beam in Sec.~\ref{sec: a aM}, respectively.  The difference comes from switching the final-state meson with the initial-state lepton or photon. The argument for the factorization works in essentially the same way, with only a slight change due to the meson being in the initial state instead of final state. In reality, only charged light meson beams such as $\pi^{\pm}$ or $K^{\pm}$ are readily accessible in experiments, so we will consider only those beams. Then charge conservation implies a flavor change of the diffractive hadron, \ie, $h' \neq h$, which forbids the $\gamma^*$-mediated $n = 1$ channel. Therefore, the leading-power contributions to the amplitudes in Eqs.~\eqref{eq:M ll} and \eqref{eq:M aa} start with the $n = 2$ channels, which are factorized into the GPDs associated with the hadron transition $h \to h'$, as in~\cite{Berger:2001zn, Qiu:2022bpq}.

For the process in \eq{eq:M ll}, at the lowest order in QED, the high-$q_T$ lepton pair is produced via a timelike photon $\gamma_{ll}^*$ with a high virtuality $Q \sim \order{q_T}$, when it is sufficiently away from the resonance region of a heavy quarkonium. It is this highly virtual photon that couples directly to the parton lines from the $h$-$M_B$ interaction, whose virtuality $Q$ provides the hard scale that localizes the parton interactions. This is sufficient for the factorization argument. Furthermore, due to the lack of $\gamma^*$-mediated $n=1$ subprocess, the requirement of the high invariant mass for the lepton-pair is a sufficient condition for factorization, allowing us to release the high $q_T$ requirement, which is contrary to the requirement for the lepton-pair production in the SDHEP with a photon beam, as discussed in Sec.~\ref{sec:a ll}.

In contrast, the process in \eq{eq:M aa} has the two final-state photons directly couple to the parton lines, and the hard scale is solely provided by their high transverse momentum $q_T$, which is both the sufficient and necessary condition for collinear factorization. In the low-$q_T$ regime, one starts to have two widely separated scales in the same process, $q_T^2\ll \hat{s} = (p-p'+p_2)^2$, just as the photoproduction of diphoton process in Sec.~\ref{sec:a ll}, the factorization for which needs further study.

\subsection{Real photon and light meson pair production: $(CD) = (\gamma M_D)$}

Now we consider the process
\beq[eq:M aM]
	h(p) + M_B(p_2) \to h'(p') + \gamma(q_1) + M_D(q_2),
\eeq
which differs from the photoproduction of a meson pair process in Sec.~\ref{sec:a MM} by switching the initial-state photon with one of the final-state mesons. The $n = 1$ channel corresponds to the subprocess $\gamma^*(p_1) + M_B(p_2) \to \gamma(q_1) + M_D(q_2)$, which is the crossing process of meson pair production process in \paper{Brodsky:1981rp} or the meson-meson annihilation process in \paper{Qiu:2022bpq}. Depending on the quantum numbers of $M_B$ and $M_D$, this channel may or may not be present. The amplitude can be factorized into the DAs of $M_B$ and $M_D$, which can be easily generalized from the treatment in \paper{Qiu:2022bpq}.

The amplitude of $n = 2$ channel can be factorized into a GPD and two DAs, whose proof can be adapted from Sec.~\ref{sec:a MM} with straightforward modifications: one can first factorize the $D$-collinear subgraph and the soft gluons attached to it, and then do the same thing for $B$, which is sufficient to complete the proof. 

\begin{figure}[htbp]
\centering
	\begin{tabular}{cc}
		\includegraphics[scale=0.6]{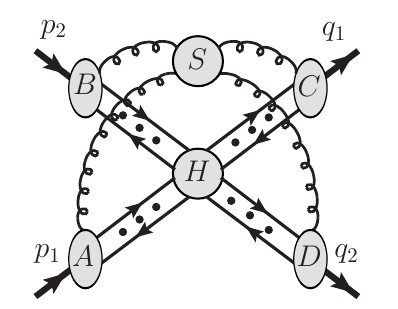}	&
		\includegraphics[scale=0.6]{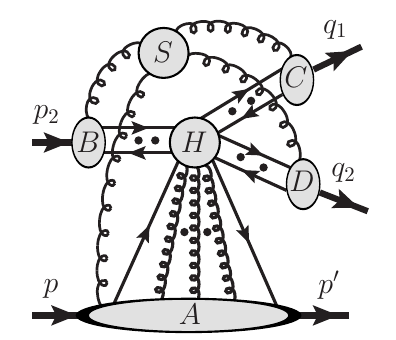}	\\
		(a) & (b) 
	\end{tabular}
\caption{Leading-region graphs for the meson-production of two mesons for (a) the corresponding $2\to2$ hard exclusive scattering~[Eq.~\eqref{eq:MM2MM}], and (b) the full SDHEP~[Eq.~\eqref{eq:M2MM}]. There can be any numbers of soft gluons connecting $S$ to each collinear subgraph. Depending on the quantum numbers, the quark lines may be replaced by transversely polarized gluon lines. The dots represent arbitrary numbers of longitudinally polarized collinear gluons.}
\label{fig:p4 leading}
\end{figure}

\subsection{Light meson pair production: $(CD) = (M_C M_D)$}
\label{subsec:M2MM}

Now we consider the process
\beq[eq:M2MM]
	h(p) + M_B(p_2) \to h'(p') + M_C(q_1) + M_D(q_2),
\eeq
whose corresponding $2\to2$ hard scattering is
\beq[eq:MM2MM]
	M_A(p_1) + M_B(p_2) \to M_C(q_1) + M_D(q_2).
\eeq
The $n = 1$ channel, $\gamma^*(p_1) + M_B(p_2) \to M_C(q_1) + M_D(q_2)$,
which may or may not contribute depending on the quantum numbers, can be analyzed in the same way that was used for analyzing the $2\to2$ hard exclusive scattering of the meson pair production in the SDHEP with a photon beam in Sec.~\ref{sec:a MM2}.
The $n = 2$ channel has leading regions shown in \fig{fig:p4 leading}, under the assumptions of strong soft-end suppression 
{\it and} a single hard scattering in which all the parton lines are off shell by the hard scale. 
Compared to the meson pair photoproduction process in Sec.~\ref{sec:a MM}, there is one more collinear subgraph in the initial state, and factorization works with a simple generalization. In both Figs.~\ref{fig:p4 leading}(a) and~\ref{fig:p4 leading}(b) one does not deform the contours of soft gluon momenta $k_s$ for their components $k_s \cdot w_A$ when they flow in the $A$-collinear subgraph. We first factorize $C$, $D$, and $B$ from $H$ sequentially, together with the soft gluons attached to them, and then group the soft gluons into the $A$-collinear subgraph to complete the proof in a way similar to what we did in Sec.~\ref{sec:a MM}.  Consequently, the amplitude of the diffractive process in \eq{eq:M2MM} can be factorized into the GPD and DAs,
\begin{align}\label{eq:p4 factorize}
	&\mathcal{M}^{(2)}_{h M_B \to h' M_C M_D} = \sum_{i, j, k, l} \int_{-1}^1 \dd{x} \int_0^1 \dd{z_B} \dd{z_C} \dd{z_D}  \nn\\
	&\hspace{1em}\times
		F^{hh'}_i(x, \xi, t) \, \phi_{j/B}(z_B) \, C_{i j \to kl}(x, \xi; z_B, z_C, z_D; q_T) \nn\\
	&\hspace{1em}\times
		\phi_{k/C}(z_C) \, \phi_{l/D}(z_D) ,
\end{align}
up to $1/q_T$ power suppressed terms, where the symbols have the same definitions as those in \eq{eq:m3factorize}, and the hard coefficient $C_{i j \to kl}(x, \xi; z_B, z_C, z_D; q_T)$ can be calculated as the scattering of two collinear parton pairs $i$ and $j$ into another two pairs $k$ and $l$. 

\subsection{Virtual photon or heavy quarkonium production}
For similar reasons as in Secs.~\ref{subsec:e virtual photon} and \ref{subsec:a virtual photon}, factorization for the diphoton production holds when one of the final-state real photons is changed to a virtual photon, which then decays into a charged lepton pair, whose total transverse momentum is balanced by the other observed real photon. Similar factorization applies to the case when both photons are virtual and decay into lepton pairs. However, it would be an experimental challenge to determine which pair of leptons comes from the same virtual photon. 

On the other hand, if the real photon in the photon-meson pair production is changed to a virtual photon, the kinematic signal is $l^+ l^- M_D$ in the final state, which has no mixed microscopic subprocesses and whose factorization is easily generalized. However, as discussed in Secs.~\ref{subsec:e virtual photon} and \ref{subsec:a virtual photon}, the heavy quarkonium production is not straightforwardly adapted while the $q_T \gg m_Q$ case is still factorizable into a heavy quarkonium DA.

\subsection{Sensitivity to the $x$-dependence of GPDs}
\label{eq:m x}

For the dilepton production process, the hard scale $q_T$ is not generated from the interaction with the partons, but instead, via the decay of a virtual photon of invariant mass $Q\ge 2q_T$ which couples to the parton lines via a single vertex. The hard scale $Q$ thus interferes with the parton momentum flows via a pointlike interaction. Hence, in the hard coefficient, the $x$-dependence factorizes from the $q_T$ dependence, which leads to the moment-type sensitivity~\cite{Berger:2001zn}. In contrast, for the diphoton production process, the hard scale $q_T$ is generated from the parton interactions, and the $q_T$ and parton momentum $x$ have to flow through the same gluon propagator in a way that they cannot be disentangled~\cite{Qiu:2022bpq}. The process can thus provide enhanced sensitivity to the $x$-dependence of GPDs.

In addition, by choosing a charged meson beam, we can eliminate the $\gamma^*$-mediated subprocesses for the dilepton or diphoton production processes.

For the photon-meson or meson-meson pair production, which have not been studied in the literature, we expect a similar kind of enhanced sensitivity to the $x$-dependence of GPDs like the diphoton production processes.

\section{Discussion}
\label{sec:discussion}

In this section, we give a few general remarks on the properties of SDHEPs, and their factorizability and sensitivities for extracting GPDs.

\subsection{Two-stage paradigm and factorization}
\label{sec:discussion two stage}
We have presented the arguments to prove the factorization of SDHEPs with different colliding beams and different types of final-state particles. 
Our proofs follow a unified two-stage approach by taking advantage of the unique feature of SDHEPs, which can be effectively separated into two stages, as specified in Eqs.~\eqref{eq:diffractive} and~\eqref{eq:hard 2to2}. By requiring $q_T\gg \rt$, we effectively force the exchanged state $A^*$ between the single diffractive transition of $h\to h'$ and the hard exclusive $2\to 2$ scattering to be a low-mass and long-lived state in comparison to the timescale~$\sim {\cal O}(1/q_T)$ of the hard exclusive process, and effectively reduce the SDHEP into two stages: single diffractive (SD) + hard exclusive (HE) with the quantum interference between these two subprocesses suppressed by powers of $\rt/q_T$. As emphasized earlier, requiring large transverse momenta for the final-state particles $C$ and $D$ is not equivalent to requiring a large invariant mass of them, $m_{CD}\gg \rt$; the latter does not necessarily guarantee a hard collision.

This two-stage paradigm gives a unified picture for the microscopic mechanism of the SDHEPs, described in \eq{eq:channels} and \fig{fig:decomposition}. It accounts for the $\gamma^*$-mediated $n=1$ channel in a coherent framework, which is usually regarded as a ``byproduct" of the GPD channel in the literature and can be easily forgotten but which is in fact one power higher than the GPD channel and should be incorporated unless it is forbidden by some quantum number conservation. 

Furthermore, this two-stage paradigm leads to a simple methodology for proving factorization of the SDHEPs in \eq{eq:sdp}, in particular, for the $n = 2$ channel. By treating the long-lived exchanged state $A^*$ as a ``meson'' capturing the quantum number of $h\to h'$ transition, we make the corresponding scattering $A^*+B\to C+D$ effectively a $2\to 2$ exclusive process with a single hard scale, whose factorization is relatively easier to prove. In this way, the factorization proof of the SDHEP can focus on its differences from the $2\to 2$ hard exclusive process.

\begin{figure}[htbp]
\centering
	\includegraphics[scale=0.6]{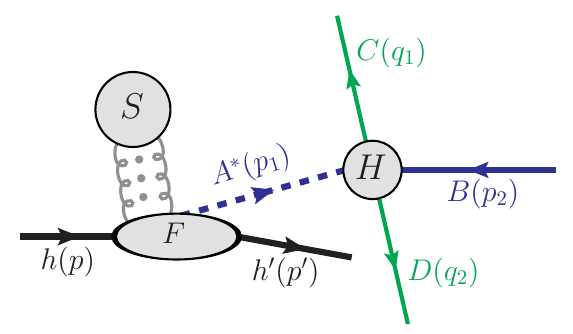}
\caption{The result of soft cancellation in the diagram \fig{fig:sdhep}(b). The cancellation of the soft gluons in the $2\to 2$ hard exclusive scattering implies the same cancellation of the soft gluons that couple to $B$, $C$, and/or $D$.}
\label{fig:sdhep soft A}
\end{figure}

The only difference between the factorization of the $2\to 2$ hard exclusive process and the full SDHEP is that the GPD channel supports both ERBL and DGLAP regions, and a Glauber pinch can exist for the DGLAP region. However, since we only have one diffractive hadron, only one component $k_s\cdot w_A$ of the soft gluon momentum $k_s$ is pinched in the Glauber region. The factorizability of the corresponding $2\to 2$ exclusive process implies that soft gluons coupling to $B$, $C$, and/or $D$ are canceled, which applies equally to the situation of SDHEPs. The rest of the soft gluons only couple to the diffracted hadron and can be grouped into the collinear subgraph of the diffractive hadron $h\to h'$; see \fig{fig:sdhep soft A} as an illustration. The factorization of soft gluons leads to the independence among different collinear subgraphs, and help to establish the factorization of the collinear subgraph of the diffractive hadron into a universal GPD, and the other collinear subgraphs into universal meson DAs.

\subsection{Assumptions for the exclusive factorization}
\label{subsec:assumption}

As discussed in \paper{Qiu:2022bpq}, the keys to collinear factorization are the cancellation of soft subgraphs that connect to different collinear subgraphs and the factorization of all collinear subgraphs from the infrared-safe short-distance hard part.

The first assumption that we made is that the leading active quark lines or transversely polarized gluon lines from the mesons must be coupled to the hard interaction, but not to the soft subgraph, for which we effectively assume that we could get an additional suppression from the expected end point behavior of meson wave function, when one of the active quarks (or gluons) has a soft momentum, which we have referred to as the soft-end suppression. The result of this assumption is that, to the leading power, the soft subgraph is only connected to collinear subgraphs by gluon lines that are longitudinally polarized, for which Ward identity can be applied to factorize them onto Wilson lines. The soft Wilson lines are only connected to the rest of the graph by colors, and can be disentangled and factorized from the collinear subgraphs because the collinear subgraphs are in color singlet states, which is an important feature of exclusive processes. Consequently, the soft cancellation for the factorization of SDHEPs is very different from typical soft cancellation for the factorization of inclusive processes~\cite{Collins:1989gx}.

Another consequence of the soft-end suppression is that we are allowed to constrain the light-cone parton momenta of the mesons on the real axis and arrive at a definition of meson DA, $\phi(z)$ with $0 < z < 1$, as argued at the end of Sec.~\ref{sec:dvcs}.

This assumption was also applied to the most factorizations of exclusive processes involving high-momentum mesons, notably for the pion form factor and large-angle production processes; see the review~\cite{Brodsky:1989pv}. Even though the soft-end region was conjectured to be Sudakov suppressed in~\cite{Brodsky:1989pv}, which is more than the power suppression taken as our assumption, a more extensive discussion on this issue is still lacking in the literature. We wish to come back to this in the future.

The second assumption that we implicitly made is that there is only one single hard interaction in which all the parton lines are effectively off shell by the hard scale. This applies especially to the meson-production of a meson pair process in Sec.~\ref{subsec:M2MM}. It is well known that the exclusive hadron-hadron scattering into large-angle hadrons can happen via multiple hard interactions, which has an enhanced power counting with respect to the single hard interaction~\cite{Landshoff:1974ew, Botts:1989kf}. We have shown the factorization for the hard exclusive $2\to 2$ meson-meson scattering and the corresponding SDHEP with a meson beam for the single hard interaction case. Within the two-stage paradigm, it is unclear to us whether the factorization of the large-angle meson-meson scattering via multiple hard interactions can imply a corresponding factorization for the SDHEP with a meson beam; we leave that for future study.

One may also consider representing $A^*$ as a sum over virtual hadronic states, instead of the expansion in terms of partonic states like $[q\bar{q}']$ and $[gg]$. 
However, the exchanged state $A^*$ in the SDHEP enters a hard collision, which has a resolution scale $1/Q$ much smaller than the typical hadronic scale, and therefore it is the partonic degrees of freedom inside the virtual hadronic state or the diffractive hadron that are probed. 
For example, 
the leading-power contribution from a virtual hadronic state should also be mediated by two active parton lines, just as in Figs.~\ref{fig:dvcs meson}, \ref{fig:m2 leading}(a), \ref{fig:a MM leading}(a), and~\ref{fig:p4 leading}(a), along with the same short-distance hard part as the $n=2$ partonic channel in connection with GPDs.
In principle, to this power, one should add all the two-parton-mediated contributions from all possible virtual hadronic states of the same diffractive hadron, which could possibly recover the full contributions from the corresponding GPDs of the same hadron, but, only from their ERBL region. GPDs also contain the DGLAP region, which cannot be covered by the subprocesses mediated by virtual hadronic states.
The approach of taking out a virtual meson $A^*$ from the $h\to h'$ transition, described by some form factor $F^{A}_{h\to h'}(t)$, followed by extracting two parton lines via its distribution amplitude, should also be captured by the GPD of $h\to h'$ transition in a more general sense.
The choice to represent $A^*$ by a single virtual meson state, like the Sullivan process, is therefore an additional approximation. 
On the other hand, the expansion in terms of the number of partons, $n$, is an expansion in powers of $1/Q$.

\subsection{Why single diffractive?}

From the procedure for proving factorization in the two-stage paradigm, it is easy to understand the importance of the {\it single} diffraction for factorizability of the exclusive process. The whole difficulty from the diffraction is the DGLAP region that pinches one component of the soft gluon momentum in the Glauber region, and we get away with it by only deforming the other components associated with other mesons. After factorizing out all the other mesons, the rest of the soft gluons are only coupled to the diffracted hadron and can be grouped together into this hadron's GPDs.

\begin{figure}[htbp]
\centering
		\includegraphics[scale=0.6]{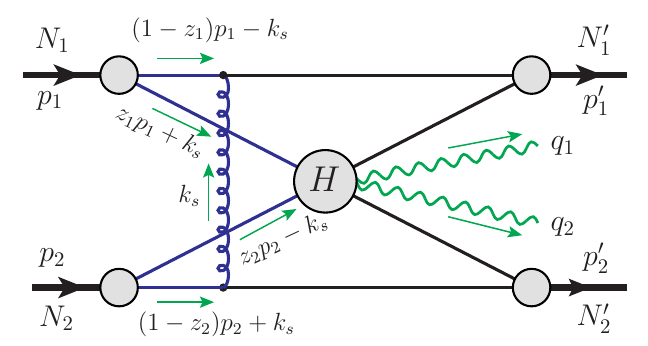}
\caption{Diphoton production in a double diffractive hard exclusive scattering process between two head-on hadrons $N_1$ and $N_2$ along the $z$ axis.}
\label{fig:double diffractive}
\end{figure}

If we consider the double diffractive process, as shown in \fig{fig:double diffractive}, the soft gluon $k_s$ exchanged between the remnants along opposite directions is pinched in the Glauber region for both $k_s^+$ and $k_s^-$, and thus no deformation can be done to get it out. As a result, this process cannot be factorized, even if we do have a hard scale provided by the transverse momentum $q_T$ of the final-state photon pair.

Similar conclusion holds for the {\it inclusive} diffractive processes~\cite{Soper:1997gj, Collins:1997sr}. The observation of the diffracted hadron anchors the inclusive sum over the final state and forbids the use of unitarity to cancel the Glauber gluon exchanges. While the soft gluon momentum can be deformed out of the Glauber region for single diffractive inclusive processes~\cite{Collins:1997sr}, in a similar way to the exclusive processes discussed in this paper, it does not work for inclusive diffractive hadron-hadron scattering~\cite{Landshoff:1971zu, Henyey:1974zs, Cardy:1974vq, DeTar:1974vx, Collins:1992cv, Soper:1997gj}.

This phenomenon is very similar to the factorization of Drell-Yan process at high twists~\cite{Qiu:1990xxa, Qiu:1990xy}, where the hadron connected by more than two active partons to the hard part is analogous to the diffracted hadron here. Even though the extra transversely polarized gluon lines at a high twist may be confused by soft gluons and endangers factorization, this is still factorizable as one can first factorize soft gluons out of the other hadron at the leading twist, similar to the procedure for the single diffractive process here that we first factorize the soft gluons out of the other mesons. This can only be done at the {\it first} subleading twist for which one of the two hadrons still has a twist-2 PDF involved, and so the Drell-Yan process is not factorizable beyond the first nonvanishing subleading twist, similar to the nonfactorizability of double diffractive processes.

\subsection{Comparison to high-twist inclusive processes}

\begin{figure}[htbp]
\centering
	\begin{tabular}{cc}
		\includegraphics[scale=0.59]{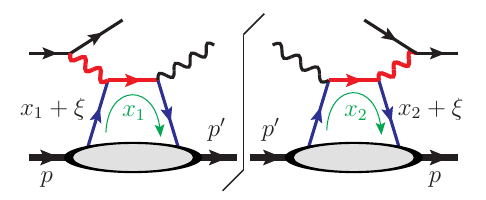}	&
		\includegraphics[scale=0.59]{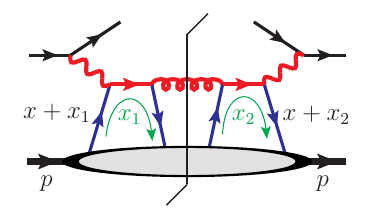}	\\
		(a) & (b) 
	\end{tabular}
\caption{Sample leading-order cut diagrams for (a) DVCS amplitude squared and (b) inclusive DIS cross section at twist-4. The red thick lines indicate the hard parts, and the blue lines are collinear partons.}
\label{fig:t4}
\end{figure}

\begin{figure}[htbp]
\centering
	\begin{tabular}{cc}
		\includegraphics[scale=0.6]{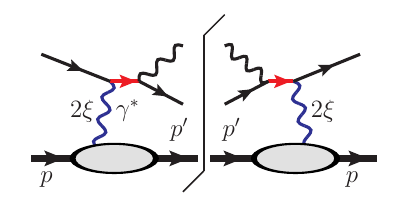}	&
		\includegraphics[scale=0.6]{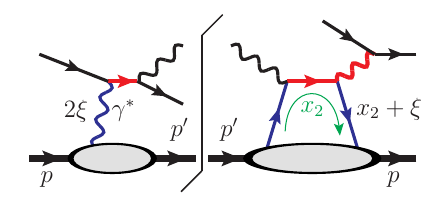}	\\
		(a) & (b) 
	\end{tabular}
\caption{Sample cut diagrams of the amplitude squared of the real photon electroproduction process for (a) the $\gamma^*$-mediated channel, and (b) the interference between the $\gamma^*$-mediated channel and GPD channel. The red thick lines indicate the hard parts, and the blue lines are collinear partons or photons.}
\label{fig:t23}
\end{figure}

The factorization of exclusive processes at the amplitude level shares many common features with the inclusive process factorization at a high twist. Taking the leading-order DVCS amplitude as an example, we show the amplitude square as a cut diagram in \fig{fig:t4}(a), which is compared with one of the leading-order diagrams of the inclusive DIS at twist-4 in \fig{fig:t4}(b). They only differ in that the cut line for the DVCS forces an exclusive final state. Both diagrams have two collinear parton lines connecting the hadron-collinear subgraph to the hard part, in both the amplitude to the left of the cut and conjugate amplitude to the right. In this sense, the DVCS amplitude squared corresponds to a twist-4 contribution to the cross section of the real photon electroproduction process. On the other hand, the amplitude squared of the $n = 1$ channel for the $\gamma^*$-mediated subprocess corresponds to a twist-2 contribution [see \fig{fig:t23}(a)], and the interference between the $n = 1$ and $n = 2$ channels corresponds to a twist-3 contribution [see \fig{fig:t23}(b)].

In the DVCS amplitude in \fig{fig:t4}(a), the two partons carry momenta $(x_1 + \xi) P^+$ and $(x_1 - \xi) P^+$ (following the directions indicated by the curved arrow), with $x_1$ integrated in $[-1, 1]$. In its conjugate amplitude, the two partons carry momenta $(x_2 \pm \xi) P^+$ with $x_2$ integrated in the same range. Similarly, for the twist-4 DIS diagram in \fig{fig:t4}(b), the amplitude part has two collinear partons with momenta $(x+x_1)p^+$ and $x_1 \, p^+$, with $x_1$ integrated in $[-1, 1-x]$. The conjugate amplitude part has two collinear partons with momenta $(x+x_2)p^+$ and $x_2 \, p^+$, with $x_2$ integrated in the same range. In both cases, the $x_1$ and $x_2$ integrations are not related and to be integrated independently. Only the total momentum of the two partons, which is $2\xi P^+$ for the DVCS and $x p^+$ for the twist-4 DIS, is observable, whose dependence is probed by the experiment.

On the other hand, there are soft breakpoint poles of $x_1$ (or $x_2$), given by the situations when one of the two partons has zero momentum, which is $x_1 = \pm \xi$ for DVCS and $x_1 = 0$ or $-x$ for twist-4 DIS. However, those poles are not pinched and they happen at the middle part of the $x_1$ integration range. As a result, we can deform the contour of $x_1$ to avoid them, just as discussed around \eq{eq:principle value}. This situation is contrary to the DA factorization, for which the soft poles happen at the end points of the DA integration and cannot be deformed away, which requires us to make the soft-end suppression assumption in Sec.~\ref{subsec:assumption}.

\subsection{Angular correlation}
\label{subsec:angular}

\begin{figure}[htbp]
\centering
		\includegraphics[scale=0.5]{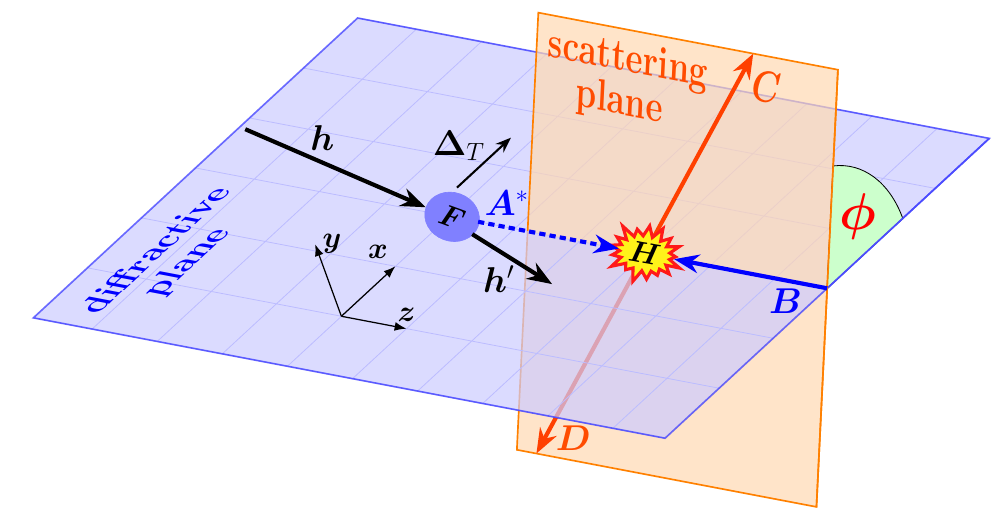}
\caption{The frame to study the SDHEP is the c.m. frame of $A^*$ and $B$. $F$ denotes the (nonperturbative) diffraction process $h\to h' + A^*$, which happens in the blue plane (``diffraction plane"), and $H$ denotes the hard interaction between $A^*$ and $B$ to produce $C$ and $D$, which happens in the orange plane (``scattering plane"). The two planes form an angle of $\phi$ and intersect at the collision axis between $A^*$ and $B$, which is chosen as the $z$ axis. $\bm{\Delta}_T$ denotes the transverse momentum of $A^*$ in the lab frame, along which the $x$ axis is chosen.}
\label{fig:frame}
\end{figure}

By the two-stage paradigm, the most natural frame for the study of the SDHEP is the c.m. frame of the $A^*$ and $B$ with $A^*$ along the $z$ axis, which is shown in \fig{fig:frame}, where the diffraction process [\eq{eq:diffractive}] happens in the blue plane, and the hard scattering process [\eq{eq:hard 2to2}] happens in the orange plane. The $x$ axis lies on the diffraction plane and points to the same direction as $\bm{p}_{1T} = \bm{\Delta}_T \equiv \bm{p}_T - \bm{p}'_T$ in the lab frame, as shown in \fig{fig:frame}. This frame can be obtained from the lab frame, the c.m. frame of the colliding beams of $h$ and $B$, by boosting along $-\bm{p}'$, as defined in \paper{Berger:2001xd}.

Each event of SDHEP can be described by five independent kinematic variables: the transverse momentum $q_T$ (or equivalently, its polar angle $\theta$) of one of the two back-to-back final-state particles ($C$ or $D$) in their c.m. frame, which is our hard scale, the azimuthal angle variable $\phi$ of this particle in the same frame, which is directly connected to the angle between the diffractive plane and the scattering plane, the c.m. energy squared $\hat{s}$ of the hard collision between $A^*$ and $B$, and the transverse momentum shift $\bm{\Delta}_T$ of the diffracted hadron in the lab frame.
They can be equivalently transformed into $(\theta, \phi, \xi, t, \phi_{\Delta})$, where $\phi_{\Delta}$ is the azimuthal angle of $\bm{\Delta}_T$ in the lab frame.
The distribution of $\phi_{\Delta}$ is determined by the diffraction process, in particular, by the spin state of the initial state hadron $h$.

The angle $\phi$ describes the angular correlation between the diffraction and the hard collision. Its distribution is solely determined by the spin states of $A^*$ and $B$. If we denote the helicities of $A^*$ and $B$ by $\lambda_A$ and $\lambda_B$, respectively, then the $\phi$ dependence of the hard scattering amplitude is captured by a phase factor
\beq
	e^{i(\lambda_A - \lambda_B) \phi}.
\eeq

For the $n = 1$ channel, $A^* = \gamma^*$ has three helicity states $(+1, 0, -1)$. For the $n = 2$ channel, the quark GPDs have three possible helicities $\lambda_A^q = 0$ or $\pm 1$, where $\lambda_A^q = 0$ has two independent contributions from the unpolarized and polarized GPDs, while $\lambda_A^q = \pm 1$ is given by the two transversity GPDs. Similarly, the gluon GPDs also have three helicities $\lambda_A^g = 0$ or $\pm 2$, with $\lambda_A^g = 0$ receiving contributions from both the unpolarized and polarized GPDs and $\lambda_A^g = \pm 2$ from the two transversity GPDs.

The interference between $(\lambda_A, \lambda_B)$ and $(\lambda_A^{\prime}, \lambda_B^{\prime})$ leads to the azimuthal correlations
\beq
	\cos(\Delta\lambda_A - \Delta\lambda_B)\phi,\;
	\mbox{ and/or }\;
	\sin(\Delta\lambda_A - \Delta\lambda_B)\phi,
\eeq
depending on details of the interaction, where $\Delta\lambda_{A, B} = \lambda_{A, B} - \lambda_{A, B}^{\prime}$. Extracting different trigonometric components of the azimuthal distribution is a great way to disentangle different GPD contributions, in a way similar to using the angular modulations
in the semi-inclusive DIS to extract different transverse momentum dependent PDFs, or TMDs~\cite{Bacchetta:2006tn}. Similarly, the angular distribution of the lepton pair in the Drell-Yan process~\cite{Lam:1978pu} was studied to capture richer structures of QCD dynamics than the production rate alone.
Because of the exclusive nature, the SDHEP cross section can receive contributions from the interferences among any two of $A^* = \gamma^*, [q\bar{q}']$ and $[gg]$ channels as well as their different polarization states.

\subsection{Sensitivity to $x$-dependence of GPDs}
\label{sec:x-dependence}
We are considering the sensitivity to the $x$-dependence of GPDs from the tree-level hard part $C(x, Q)$, where $Q$ is the external observable not associated with the diffractive hadron.\footnote{Even though the GPD variable $\xi$ is also in the hard coefficient $C$ and is directly observable from the diffracted hadron momentum, we do not consider it to be included in $Q$, but instead it always comes with $x$ and is suppressed in $C(x, Q)$.} As discussed in the previous sections, we consider the two types of sensitivity:
\begin{enumerate}
\item[(I)] \emph{Moment-type sensitivity}: $C(x, Q)$ factorizes into an $x$-dependent part and $Q$-dependent part,
	\beq[eq:hard coeff factorize]
		C(x, Q) = G(x) \, T(Q).
	\eeq
	In this case, the measurement of the $Q$ distribution, which is fully captured by the predictable $T(Q)$, does not help in probing the $x$-dependence of GPDs, and all the sensitivity is in the moment-type quantity
	\beq
		\int_{-1}^1 \dd{x} G(x) \, F(x, \xi, t).
	\eeq
	We call a process with only moment-type sensitivity a type-I process.
\item[(II)] \emph{Enhanced sensitivity}: $C(x, Q)$ does not factorize, in the sense of \eq{eq:hard coeff factorize}. Then, the distribution of $Q$ depends on the detailed $x$ distribution in the GPD. To some extent, $Q$ is the ``conjugate variable" of $x$, and they are related in the amplitude
	\beq[eq:enhanced]
		\mathcal{M}(Q) \sim \int_{-1}^1 \dd{x} C(x, Q) \, F(x, \xi, t)
	\eeq
	through the transformation kernel $C(x, Q)$, which is, in general, not invertible, of course.
	We call a process with enhanced sensitivity a type-II process.
\end{enumerate}

Only having moment-type sensitivity is far from enough, even with next-to-leading-order hard coefficients and evolution effects included~\cite{Bertone:2021yyz}, as also confirmed in practical fits of GPDs~\cite{Diehl:2004cx, Hashamipour:2020kip, Hashamipour:2021kes, Guo:2022upw}. Given the complicated functional dependence of the GPD on $x$ plus its entanglement with $\xi$ and $t$ variables, one should have as much enhanced sensitivity as possible while having as many independent moment constraints. Among the processes that have been studied in the literature, only the DDVCS~\cite{Guidal:2002kt}, photoproduction of photon-meson pair~\cite{Boussarie:2016qop, Duplancic:2018bum}, and meson-production of diphoton~\cite{Qiu:2022bpq} processes are type-II processes, and all the other processes~\cite{Ji:1996nm, Radyushkin:1997ki, Brodsky:1994kf, Frankfurt:1995jw, Berger:2001xd, Berger:2001zn, Pedrak:2017cpp} belong to type I.

\begin{figure}[htbp]
\centering
	\begin{tabular}{cc}
		\includegraphics[trim={0 -0.3cm -0.2cm 0}, clip, scale=0.7]{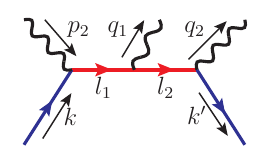}	&
		\includegraphics[trim={-0.3cm 0 0.35cm 0 0}, clip, scale=0.7]{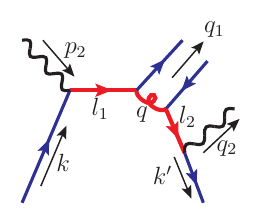}	\\
		(a) & (b) 
	\end{tabular}
\caption{Sample diagrams for the hard scattering of the single diffractive (a) photoproduction of diphoton process, and (b) photoproduction of photon-meson pair process. The red thick lines indicate the propagators in the hard part, and the blue lines are amputated parton lines that are put on shell and massless.}
\label{fig:hard}
\end{figure}

A careful examination of the denominator structure of the leading-order hard part of the partonic scattering can help understand and identify the difference in the $x$-sensitivity from these two types of processes.
The type-I processes have one common feature that every internal propagator can be made to have one end connect to two on-shell massless external lines, whether the external line is an amputated parton line or a real massless particle. Take the photoproduction of diphoton process, with one of its hard scattering diagrams in \fig{fig:hard}(a), as an example, the propagator of momentum $l_1$ is connected to an amputated parton line of on-shell momentum $k=(x+\xi)\hat{P}$ and the incoming photon line of momentum $p_2$, while the propagator of momentum $l_2$ is connected to an amputated parton line of momentum $k'=(x-\xi)\hat{P}$ and the outgoing photon line of momentum $q_2$. In the c.m. frame of the hard exclusive collision as defined in \fig{fig:frame}, we have
\begin{align}\label{eq:kin in}
	\hat{P}^{\mu} &= \pp{P^+, 0^-, \bm{0}_T}, \quad
	p_2^{\mu} = \pp{0^+, p_2^-, \bm{0}_T} ,	\nn\\
	\Delta^+ & = p_1^+ = 2\xi P^+ = p_2^- = \sqrt{\hat{s}/2}	\,,	
\end{align}
and the final-state momenta $q_1$ and $q_2$, which define the hard scale $q_T$, 
\begin{align}\label{eq:kin out}
	q_1^{\mu} & = \frac{\sqrt{\hat{s}}}{2} (1, \bm{n}) 	\nn\\
		&= \pp{ \sqrt{ \frac{\hat{s}}{2} } \frac{1 + \cos\theta}{2}, \sqrt{ \frac{\hat{s}}{2} } \frac{1 - \cos\theta}{2}, \bm{q}_T }	,	\\
	q_2^{\mu} & = \frac{\sqrt{\hat{s}}}{2} (1, -\bm{n}) 	\nn\\
		&= \pp{ \sqrt{ \frac{\hat{s}}{2} } \frac{1 - \cos\theta}{2}, \sqrt{ \frac{\hat{s}}{2} } \frac{1 + \cos\theta}{2}, -\bm{q}_T },	\nn
\end{align}
where we present them first in terms of Cartesian coordinates with $\bm{n}$ being a unit spatial vector defined as $\vec{q}_1/|\vec{q}_1|$
and then in light-front coordinates, and we also introduced the polar angle $\theta$ to represent $q_T \, ( = \sqrt{\hat{s}} \, \sin\theta / 2)$. With all external momenta defined in Eqs.~(\ref{eq:kin in}) and (\ref{eq:kin out}), we can express the virtuality of the internal momentum $l_1$ as
\beq[eq:l1]
	l_1^2 = 2 k \cdot p_2 = 2 (x+\xi) \hat{P} \cdot p_2  = \frac{x+\xi}{2\xi} \, \hat{s} \equiv x_{\xi}  \, \hat{s} \,,
\eeq
where $x_{\xi} = (x + \xi) / 2\xi$ is the same as the $z_1$ variable defined in Eq.~(3.16) of Ref.~\cite{Qiu:2022bpq}.
Similarly, we have the virtuality of the other internal momentum $l_2$ as
\beq[eq:l2]
	l_2^2 = 2 k^{\prime} \cdot q_2 = 2(x-\xi) \hat{P} \cdot q_2 
		= x_{\xi}^{\prime} \cdot \cos^2(\theta/2) \,  \hat{s} \,,
\eeq
where $x_{\xi}^{\prime} = (x - \xi) / 2\xi = x_{\xi} - 1$. 
And then the hard coefficient of the diagram \fig{fig:hard}(a) takes a factorized form, 
\begin{align}\label{eq:hard diphoton}
	 C(x, \xi, \cos\theta) 
		& \propto \frac{1}{(l_1^2 + i\varepsilon) (l_2^2 + i\varepsilon) }	\\
		& \propto \bb{ \frac{1}{(x_{\xi}+i\varepsilon)(x_{\xi}^{\prime} + i\varepsilon)} } \cdot \frac{1}{\cos^2(\theta/2)}	\, ,\nn
\end{align}
in which the dependence on $\theta$ (or equivalently, $q_T$) is factorized from the momentum fraction $x$ of the relative momentum of the active $[q\bar{q}]$ pair. 
This is an immediate consequence of having the internal propagator directly connected to two external on-shell massless particles.
Generally, as a result of connecting to two on-shell massless lines with momenta $e_1$ and $e_2$, the virtuality of the internal propagator is just a product $e_1 \cdot e_2$, which simply factorizes into a GPD-$x$ (or DA-$z$) dependent factor and a factor that depends on the external observable such as $\theta$ in \eq{eq:l2}.
This example also indicates that the poles of $x$ take place at $x_{\xi} = 0$ or $x_{\xi}^{\prime} = 0$, that is, $x = \pm \, \xi$, which are at the boundary points between the DGLAP and ERBL regions.

In contrast, a type-II process has at least one internal line in the hard part that cannot be made to have either end connect to two on-shell massless lines. We take the photoproduction of a photon-meson pair process as an example, for which one hard scattering diagram is shown in \fig{fig:hard}(b). The kinematics is the same as in Eqs.~\eqref{eq:kin in} and~\eqref{eq:kin out}, and two of the propagators, $l_1$ and $l_2$, are the same as the previous diphoton production example, given in Eqs.~\eqref{eq:l1} and~\eqref{eq:l2}.

However, the gluon propagator $q$ is connected to $l_1$ on one end and to $l_2$ one the other end, both of which are not on shell. Letting the outgoing quark line along $q_1$ have its momentum $z q_1$, we have the gluon momentum,
\beq
	q = k + p_2 - z q_1 
		= (x + \xi) \hat{P} + p_2 - z q_1	\,,
\eeq
which has the virtuality
\beq[eq:gluon aM]
	q^2 = \hat{s} \bb{ x_{\xi} \pp{1 - z \sin^2(\theta/2)} - z \cos^2(\theta/2) } \,.
\eeq
This leads to a hard coefficient that does not take a simple factorized form to separate the $(x_{\xi}, z)$ dependence from the observable $\theta$, and therefore the distribution of $\theta$ contains extra sensitivity to the shape of $x$ and $z$ in the GPD and DA, respectively.

Compared to \eq{eq:hard diphoton}, the gluon propagator in \eq{eq:gluon aM} leads to some new poles of $x$, at
\beq
	x_{\xi} = \frac{z \cos^2(\theta/2)}{1 - z \sin^2(\theta/2)} \in [0, 1],
	\;
	\mbox{ for }
	z \in [0, 1],
\eeq
which corresponds to $x\in [-\xi, \xi]$, and thus lies in the ERBL region. These are not pinched poles, so do not pose any theoretical obstacles, but are just the regions where we need to deform the contour of $x$ to avoid them. 

Similarly, in \fig{fig:hard}(a), if we make the photon $q_2$ virtual in the diphoton production process, the photon momenta in \eq{eq:kin out} will become
\begin{align}\label{eq:kin out m}
	q_1^{\mu} & = \frac{\sqrt{\hat{s}}}{2} (1-\zeta) (1, \bm{n})	\,, 	\\
	q_2^{\mu} & = \frac{\sqrt{\hat{s}}}{2} (1+\zeta, -(1-\zeta) \bm{n}) 	\,,	\nn
\end{align}
where $\zeta = Q^{\prime 2} / \hat{s}$ with $Q^{\prime 2} = q_2^2$ being the virtuality of the photon $q_2$ that decays into a lepton pair. Then the propagator $l_2$ becomes
\beq[eq:l2 virtual a a]
	l_2^2 = \hat{s} \cc{ x_{\xi}^{\prime} \, \cos^2(\theta/2) + \zeta \bb{ 1 + x_{\xi}^{\prime} \, \sin^2(\theta/2) } }	\,,
\eeq
which differs from \eq{eq:l2} by having an additional term proportional to $\zeta$ that introduces an extra scale dependence. By varying $\zeta$ and $\theta$, one can get extra sensitivity to the $x$-dependence of the GPD. This is the same mechanism that gives the enhanced $x$-sensitivity as the DDVCS process~\cite{Guidal:2002kt} which we discussed around \eq{eq:qqsq}.
This propagator [\eq{eq:l2 virtual a a}] leads to a new pole of $x$ at
\beq
	x_{\xi}^{\prime} = \frac{- \zeta}{\cos^2(\theta/2) + \zeta \sin^2(\theta/2)} \in [-1, -\zeta],
	\;
	\mbox{ for } \theta \in [0, \pi],
\eeq
that is $x \in [-\xi, (1-2\zeta)\xi] \subset [-\xi, \xi]$, which is again inside the ERBL region.

By comparison, the type-I processes are usually topologically or kinematically simpler than the type-II processes, so their theoretical analysis and hard coefficient calculations are usually easier. The type-II processes introduce enhanced sensitivity to the $x$ dependence by having extra scale dependence that entangles with the $x$ flow. For the two type-II examples we have just examined, the photon-meson pair production process differs from the DVMP process by having one extra photon attaching to the active parton lines, while the virtual photon production process differs from the real photon production process by having an extra scale $Q'$ which is in turn achieved by having that photon decay into {\it two} leptons. 
In general, extra scale dependence is introduced by more complicated topology,\footnote{Here, we consider virtual or massive particles as having more complicated topology than real massless particles, even in the case when the mass scale is not associated with virtual particle decay.} which is usually the necessary condition for enhanced sensitivity.

One important role that the SDHEP plays is that it sets a template for listing a number of processes, which we have categorized according to the beam types. We have shown the proof of factorization in a general sense. Within this framework, one shall study as many independent processes as possible, which should in turn constrain the $x$ dependence of GPDs as much as possible.

\section{Summary and outlook}
\label{sec:summary}

GPDs are important and fundamental nonperturbative parton correlation functions in QCD, and carry rich information on the spatial distributions of quarks and gluons in a confined hadron, which could provide the unprecedented and much needed information to uncover the mystery of QCD tomography. The knowledge of GPDs can help answer many unknown questions surrounding QCD and strong interaction physics.

In this paper, we proposed a general type of exclusive processes to help extract GPDs from experimental measurements, referred to as the single diffractive hard exclusive processes (SDHEPs).
The SDHEP keeps intact the hadron to be studied, and has it undergo a diffraction from momentum $p$ to $p'$, as specified in Eq.~(\ref{eq:diffractive}), which is characterized by a large momentum transfer $\Delta = p - p'$ with a small invariant mass $\sqrt{-t} = \sqrt{-\Delta^2}$.
On the one hand, the large momentum transfer guarantees a high-energy exclusive scattering with the colliding beam of leptons, photons or mesons to produce two back-to-back particles with large transverse momenta $q_T$, as defined in Eq.~(\ref{eq:hard 2to2}). 
On the other hand, the small invariant mass $\rt$ ensures that the exchanged state $A^*$ is so long lived that the ``hard probe"---the $2\to 2$ hard exclusive process does not alter the internal structure and properties of the diffracted hadron, which are quantified by the factorized GPDs.  

In this way, the SDHEP is a generic $2\to 3$ exclusive process with two distinct momentum scales: a hard scale $q_T$ that defines the resolution of the hard probe for it to see the particle nature of the quarks and gluons at a short-distance scale, and a soft scale $\rt\ll q_T$ (or $|(p-p')_T|\ll q_T$) for it to be sensitive to the long-distance partonic landscape or spatial tomography inside a confined hadron. 
The condition $\rt\ll q_T$ is necessary to suppress the quantum entanglement between the hard probe taking place at the scale of $q_T$ and the structure information at the scale of $\rt \sim |(p-p')_T|$ inside the diffracted hadron, allowing us to factorize the latter into universal and process-independent GPDs. It is the Fourier transform of GPDs' dependence on $(p-p')_T$ that provides the access to the spatial distribution of quarks and gluons inside the hadron in slices of different values of $x$.

Even though the SDHEP is the {\it minimal} configuration that can ensure the condition of $\rt\ll q_T$, we have demonstrated that it is generic enough that nearly all the processes that have been considered in the literature for extracting GPDs fit into the SDHEP framework. 
One can further generalize the SDHEP to processes with more than two large-transverse-momentum particles in the final state, which we leave for future study.

We have shown that the condition $\rt\ll q_T$ is not only {\it necessary}, but also generally {\it sufficient} for the SDHEP to be factorized into hadron GPDs, convoluted with perturbatively calculable hard matching coefficients, provided that the hard exclusive $2\to2$ scattering is dominated by a single hard scale. The proof is given under a two-stage paradigm, introduced in Eqs.~\eqref{eq:diffractive}-\eqref{eq:channels} and \fig{fig:decomposition}, which decomposes the whole SDHEP amplitude into a sum over the partonic channels connecting the single diffractive hadron and the hard exclusive $2\to2$ scattering. 
The two-stage paradigm incorporates the Bethe-Heitler process, which we call the $n = 1$ channel, naturally in the same framework. The factorizations of the $n = 2$ subprocesses into GPDs are proved by two steps: 
(1) prove the factorization of the corresponding $2 \to 2$ hard exclusive process, which is effectively a single-scale observable, and 
(2) complete the factorization of the SDHEP by examining the extra complication of the Glauber pinch. 
The necessary ability to get out of the Glauber region implies that only the single diffractive processes are factorizable, and double diffractive processes are not. This conclusion is similar to the inclusive diffractive processes. And we have also given the analogy to the factorization of inclusive processes at high twists.

From its two-stage paradigm, we introduced a natural reference frame for studying SDHEP, as defined in Sec.~\ref{subsec:angular}. 
This frame is not only a convenient one for proving factorization, but also provides new opportunities to relate the angular correlation between the ``diffractive plane'' and the ``scattering plane" to the spin of the beam particle $B$ and the spin structure of the exchanged state $A^*$, allowing the access to enhancing the sensitivity to various types of GPDs.

While two of the three variables $(x,\xi,t)$ of GPDs are directly related to the measured momenta of the diffractive hadron, $p-p'$, it is the relative momentum fraction $x$ of the two exchanged partons, $[q\bar{q}']$ or $[gg]$, between the diffractive hadron and the hard probe 
that is the most difficult one to extract from the experimental measurement, while it is the most important one to define the slices of the hadron's spatial tomography.
We systematically examined the sensitivity of various SDHEPs for extracting the $x$-dependence of GPDs.
We divided the sensitivity into two types: moment type and enhanced type, as defined in Sec.~\ref{sec:x-dependence}. 
We discussed the sensitivity of specific processes in Sec.~\ref{eq:e x} for SDHEPs with an electron beam, Sec.~\ref{eq:p x} for SDHEPs with a photon beam, and Sec.~\ref{eq:m x} for SDHEPs with a meson beam. 
A more general discussion was also given in Sec.~\ref{sec:x-dependence}. 
We argued that the requirement for enhanced sensitivity on $x$ is to have at least one internal propagator in the hard part that is not connected to two on-shell massless external lines on either of its ends, which usually requires observing more than one external particle that comes out of the hard scattering.

Given both the theoretical and experimental difficulties to unambiguously extract the $x$-dependence of GPDs, one should not only study as many independent GPD-related processes as possible, but also identify more processes that yield enhanced sensitivity to the $x$ dependence of GPDs.
With a generic factorization proof, the SDHEP can serve as a framework to identify and categorize all specific processes for the study of GPDs. In this paper, we categorized these processes in terms of the type of the beam colliding with the diffractive hadron.  
With the two-stage paradigm of the SDHEP, we are well motivated for the search of new processes for extracting GPDs, and in particular, their $x$-dependence.

\section*{Acknowledgments}

We thank G.T. Bodwin, M.~Diehl, X.-D.~Ji, K.-F.~Liu, Z.-E.~Meziani, B.~Pire, G. Sterman, and C.-P.~Yuan for helpful discussions and communications. 
This work is supported in part by the US Department of Energy (DOE) Contract No.~DE-AC05-06OR23177, under which Jefferson Science Associates, LLC operates Jefferson Lab, and within the framework of the TMD Collaboration. The work of Z.~Y. at MSU is partially supported by the U.S.~National Science Foundation under Grant No.~PHY-2013791, and  the fund from the Wu-Ki Tung endowed chair in particle physics.

\bibliographystyle{apsrev}
\bibliography{reference}

\end{document}